\documentclass[useAMS,usenatbib,usegraphicx]{mn2e}
\usepackage{graphicx,amssymb}
\usepackage{color}

\title[Messier 3 variables]{Long-term photometric monitoring of RR Lyr stars in M3 }

\author[Jurcsik et al.]{J. Jurcsik$^{1}$, G. Hajdu$^{1}$, B. Szeidl$^{1}$, K. Ol\'ah$^{1}$, J. Kelemen$^{1}$, \'A. S\'odor$^{1}$, A. Saha$^{2}$ \and
P. Mallick$^{2}$ and J. Claver$^{2}$\\
\\
$^{1}$Konkoly Observatory of the Hungarian Academy of Sciences, H-1525 Budapest PO Box 67, Hungary\\
$^{2}$National Optical Astronomy Observatories, Tucson, AZ 85726-6732, USA
}
\begin{document}

\date{Accepted 2011  Received 2011 Aug; in original form 2011 Aug}

\pagerange{\pageref{firstpage}--\pageref{lastpage}} \pubyear{2010}

\maketitle

\label{firstpage}
\begin{abstract}
The period-change behaviour of 134 RR Lyrae stars in the globular cluster Messier 3 (M3) is investigated on the $\sim120$-year time base of the photometric observations. The mean period-change rates ($\beta \approx 0.01 \textrm{d} \textrm{Myr}^{-1}$) of the subsamples of variables exhibiting the most regular behaviour are in good agreement with theoretical expectations based on Horizontal-Branch stellar evolution models. However, a large fraction of variables show period changes that contradict the evolutionary expectations. Among the 134 stars studied, the period-change behaviour of only 54 variables is regular (constant or linearly changing), slight irregularities are superimposed on the regular variations in 23 cases and the remaining 57 stars display irregular period variations. The light curve of $\sim50$ per cent of the RRab stars is not stable, i.e., these variables exhibit Blazhko modulation. The large fraction of variables with peculiar behaviour (showing light-curve modulation and/or irregular $O-C$ variation) indicate that, probably, variables with regular period changes incompatible with their evolutionary stages also could display some kind of instability of the pulsation light curve and/or period, but the available observations have not disclosed it yet. The temporal appearence of the Blazhko effect in some stars, and the 70--90 years long regular changes preceded or followed by irregular, rapid changes of the pulsation period in some cases support this hypothesis.

Accurate Fourier parameters of the light curves of the RRab variables are derived from all the available CCD data. The large, homogeneous sample of stars on the Oosterhoff I sequence enable us to investigate the characteristics of the light curves of the variables showing increasing and decreasing period changes. It is found that, at a given phase-difference value, period-increasing variables have 0.002--0.014 mag smaller amplitudes on the averages than period decreasing variables have. Also, at a given period, their phase differences tend to be smaller by 0.03--0.07 rad than the phase differences of variables with decreasing periods. The realness of the detected differences is proven by Monte Carlo simulations.
\end{abstract}
\section{Introduction}
Stellar evolution theory predicts that both slow expansions and contractions of stars within the pulsation instability strip are expected, that result in increasing and decreasing periods, respectively. Since the pioneering work of \citet{ma38} on the period changes of the RR Lyrae stars in $\omega$ Centauri, numerous studies have been carried out with a view to detect period changes connected with stellar evolution. The best targets for such investigations are globular clusters (GC), as most of them have a large RR Lyrae population, and their observations have extended hitherto over 50--100 years. The investigations, however, have revealed that, contrary to all expectations, both fast period increase and decrease, repeated irregular and/or abrupt changes may also characterize the period changes. This kind of period-change behaviour could not at all be reconciled with evolutionary models. Moreover, the observations have shown peculiar cases, too. Some stars showed regular or quasi-regular period-change behaviour for more than half a century, then suddenly, continuous or abrupt irregular changes in their period have taken place. One may ponder how long a time base would be enough to distinguish between evolutionary period changes and period changes caused by other effects. Several ideas have been propounded to interpret the non-evolutionary period changes, but none of them have been proven to be fully satisfactory yet.

Although the observed large, non-evolutionary period changes may frustrate the observer, there is a consent among researchers concerned that the mean value of the period-change rates of a large sample of RR Lyrae stars of a globular cluster reflects the trend of evolution consistently with the model calculation results of \citet{le91}.

One of the most thoroughly investigated globular cluster, as for the variable stars, is Messier 3 (as a recent study of the variables see e.g., \citet{ca05}). It contains one of the largest RR Lyrae populations. The latest investigation of the period-change behaviour of the RR Lyrae stars in M3 was made by \citet{co01}. This period-change study utilized the $O-C$ data published by \citeauthor{sz65} (\citeyear{sz65}, \citeyear{sz73}) in combination with CCD observations obtained between 1990 and 1997. The lack of published observational data from between 1962 to 1990 made the results, however, somewhat dubious for many of the variables. Since the archival, previously unpublished Konkoly photographic observations obtained in the years between 1964 and 1989 fill this gap, and the new CCD photometry from 2009 extends the baseline of the observations by a decade over a century, a fresh hope arises to get a better insight into the period-change behaviour of cluster variables. This provides the motivation for the present study.

\begin{table*}
\caption{Summary of the photographic observations of M3 variables.}
 \label{pg}
  \begin{tabular}{r@{\hspace{1mm}}lcccc}
  \hline
&Code/Reference&Exp. time&Time interval& No. of& Maximum number of\\
&&[min]&JD$-2400000$ &stars&data points\\
 \hline
1&\citet{ba13}&10--183	&13372--15161&132&50\\
2&\citet{he42}&2--20			&19479--20656&48&25\\
3&\citet{la22}&10--20				&22455--22840&129&136\\
4&\citet{mu33}&15--60		&23858--24317&159&91\\
5&\citet{sl29}&8			&24564--24642&47(30)&97\\
6&\citet{gr35}&8			&24647--24684&117&75\\
7&\citet{sz65}; \citet{sz73}&10--17&28963--37791&129&213\\
8&\citet{sc40}&8--10			&28964--28983&51&4\\
9&\citet{he42}&9--15		&29367--29341&48&44\\
10&\citet{be52}&5--7			&31965--32700&52&38\\
11&\citet{ku70}&na.	&33034--37131&37&180\\
12&\citet{ku61}&na.	&34421--37130&3&167\\
13&\citet{rs55}&$\sim	$15	&34447--34508&78&25\\
14&\citet{ca74}&5--10	&35190--35604&6&59\\
15&\citet{sz65}; \citet{sz73}&12--16&35920--35933&128&17\\
16&Konkoly 60-cm (present paper)&12--15			&38502,  39944&113&23\\
17&Konkoly RCC (present paper)&10--20	&42838--47555&133&168\\
\hline
\end{tabular}
\end{table*}
\begin{table*}
\caption{Summary of the CCD observations of M3 variables.}
 \label{ccd}
  \begin{tabular}{r@{\hspace{1mm}}llcccc}
  \hline
&Code/Reference&Abbreviation&Filter&Time interval&No. of&No. of\\
&&&&JD$-2400000$ &stars&data points$^*$\\
\hline
18&\citet{ca98}&C98&\textbf{\emph{B}}\textit{VI}&47954--48659&60		&63	\\
19&\citet{co01}&C01&\textbf{\emph{B}}\textit{V} &48755--50622&207	&187\\
20&\citet{K98}&K98&\textbf{\emph{V}}			 &50162--50176&42		&175\\
21&\citet{H05}&H05&\textbf{\emph{B}}\textit{VI}&50920--50965&170	&70		\\
22&\citet{be06}&B06&\textbf{\emph{B}}\textit{VI}&50896--51283&226	&205\\
23&Kitt Peak 0.9-m (present paper)&KP99&\textit{uv}\textbf{\emph{g}}\textit{ri}&51283--51287&39	&66\\
24&Konkoly Schmidt (present paper)&KS09&\textbf{\emph{V}}\textit{I} &54869--54965&126&		760\\
\hline
\multicolumn{6}{l}{\footnotesize{$^{*}$ Maximum number of frames obtained using the filter set given in boldface in the third column.}}\\
\end{tabular}
\end{table*}

\section{Data and method of investigation}

Since the pioneering work of \citet{ba13}, many photometric studies investigated the properties of the variable star population of M3. As our aim is to carry out a study of period-change behaviour of the RR Lyrae stars in M3 as complete as possible, all the measurable variables numbered from V1 to V203 have been on our list. Unfortunately, a number of stars, especially those discovered by \citet{mu33} and \citet{gr35} are situated in the  crowded, central region and/or have close, bright companion(s), therefore, their brightness measurements are unreliable and useless in most cases. Variables numbered over V204 are too sparsely (if at all) observed over the past fifty years for reliable period changes to be determined. Altogether, we have found 134 RR Lyrae stars with extended enough data for a period-change analysis.

\subsection{Previous photometric studies}

Tables \ref{pg} and \ref{ccd} summarize the photographic and CCD observations utilized in the present study. The code/reference of the photometric data, the exposuse time/filters, the time interval of the observations, the number of variables observed and the maximum number of data points in the different datasets are given in the tables.

The brightness estimates of 17 stars published by \citet{sl29}, which cannot be transformed to the magnitude scale, and the $pg_{vis}$ observations of \citet{rs55} are not used in our analysis. These data are coincidental or very close in time to other observations, thus their omission has no effect on the results. Instead of light-curve data, only timings of the mid-point on the rising branch or times of maxima ($O-C$ values) have been published by \citet{ma42}, \citet{kh66} and Meinunger \citep{we95}. These data cannot be utilized in our analysis.

The datasets obtained using the 60-cm telescope of the Konkoly Observatory, and the 1-m telescope of the Hamburg Observatory published by \citeauthor{sz65} (\citeyear{sz65}, \citeyear{sz73}) are treated separately and are listed as datasets 7 and 15 in Table \ref{pg}. 

The CCD era of the observations of M3 began in 1990 \citep{ca98} and continued with the extended measurements of \citet{co01} from 1992, 1993 and 1997, \citet{K98} from 1996, \citet{H05} from 1998, and \citet{be06} from 1998 and 1999. CCD datasets are referred by the abbreviations given in the second column of Table ~\ref{ccd}, hereafter.

\subsection{Observations published in this paper}

Photographic observations of M3 with the telescopes of the Konkoly Observatory were obtained  after 1962, which data have not been published previously. The exposures taken without filter  with the 60-cm telescope on two nights in 1964 and 1968, and with the 1-m RCC telescope between 1976--1980 and 1988--1989 using  103aO plates are processed in the present work.

The photographic plates were digitalized using an Umax PowerLook 3000 flatbed transparency scanner. The photographic densities of the variables and comparison stars were measured using standard aperture photometry packages of IRAF\footnote{{\sc IRAF} is distributed by the National Optical Astronomy Observatories, which are operated by the Association of Universities for Research in Astronomy, Inc., under cooperative agreement with the National Science Foundation.}. The brightness of the variables were measured using comparison stars selected from the photographic standard sequence of \citet{sa53}.

CCD photometric data were obtained with the 0.9-m telescope of the Kitt Peak National Observatory on 15--19 April
1999 using the Cassegrain imaging camera and the "T2KA" Tektronix detector with 2K$\times$2K (24-$\micron$ pixels) and the extended Thuan--Gunn $uvgri$ filter set.  The detector offered a field of view of approximately 24 arcmin square. Data were reduced using standard IRAF processing procedures and photometric magnitudes were measured using PSF fitting techniques. From the 40 light curves evaluated, 39 are utilized as V161 does not have long enough  photometric record to analyse its period changes.

CCD observations of M3 were performed with the 60/90/180-cm Schmidt telescope at the Piszk\'estet\H{o} Mountain Station of the Konkoly Observatory using a Photometrics camera with a Kodak KAF-1600 1024 $\times$ 1536 chip between 6 February and 13 May in 2009. The field of view was 19' $\times$ 28' with $1.0$ pixel per arcsec resolution. The observations comprised four one-week long observing runs.  In total, $\sim800$ \textit{V} and $\sim700$ $I_C$ frames were obtained.

The frames were reduced using standard IRAF packages. The reduction process included overscan, bias, dark, flat and badpixel corrections. The primary goal of the CCD observations was to extend the time base for period-change studies, therefore, aperture photometry for variables with long enough historical record was obtained, in comparison with the mean ensemble magnitudes of 23 standard stars from the list of \citet{st00}. The measurements were transformed  to the  standard Johnson--Cousins $VI_C$ system. The typical error of the data of single variables is $\sim0.02$ mag. The photometry provides the most densely covered $\sim100$ days long, accurate observations for these objects, but it is seriously biased in amplitude and mean magnitude for crowded stars.

The above listed photographic and CCD time series of the variables are available from the electronic edition of this article. Table~\ref{data} gives a sample of the data regarding its form and content. 

\begin{table}
\caption{Photographic and CCD observations of M3 variables published in this paper. The HJDs represent the times of mid-exposure. The complete table is given in the electronic version of the article as supporting information.}
\label{data}
\begin{tabular}{ll@{\hspace{5mm}}l@{\hspace{7mm}}l@{\hspace{7mm}}l}
\hline
Star&HJD$-2400000$&mag&band&source$^a$\\
\hline
V9&38502.402&15.92&$pg$&B\\
.	&....	&....	&....\\
V9&42838.586&15.85&$pg$&P\\
.	&....	&....	&....\\
V9&51283.784&16.26&$u$&K\\
.	&....	&....	&....\\
V9&51283.7782&16.11&$v$&K\\
.	&....	&....	&....\\
V9&51283.7143&15.95&$g$&K\\
.	&....	&....	&....\\
.	&....	&....	&....\\
V9&54869.5347&15.988&$V$&S\\
.	&....	&....	&....\\
V9&54872.4294&15.052&$I$&S\\
.	&....	&....	&....\\
\hline
\multicolumn{4}{l}{$^a$ B: Budapest 60-cm telescope, photographic}\\
\multicolumn{4}{l}{\hspace{1.7mm} P: Piszk\'estet\H{o} 1-m telescope, photographic}\\
\multicolumn{4}{l}{\hspace{1.7mm} K: Kitt Peak 0.9-m telescope, CCD}\\
\multicolumn{4}{l}{\hspace{1.7mm}  S: Piszk\'estet\H{o} 60/90-cm Schmidt telescope, CCD}\\
\hline
\end{tabular}
\end{table}

\begin{table*}
\caption{Summary of datasets of the variables used in the period-change analysis. The available observations are marked by `+' symbol. Asterisks indicate the datasets used for the construction of the mean light curves. The columns refer to the datasets' number as given in Tables~\ref{pg} and \ref{ccd}.  The complete table is available in the electronic version of the article as supporting information.}
\begin{minipage}{\textwidth}
\label{felhasznalt}
\begin{tabular}{l@{\hspace{2mm}}c@{\hspace{3.7mm}}c@{\hspace{3.7mm}}c@{\hspace{3.7mm}}c@{\hspace{3.7mm}}c@{\hspace{3.7mm}}c@{\hspace{3.7mm}}c@{\hspace{3.7mm}}c@{\hspace{3.7mm}}c@{\hspace{3.7mm}}c@{\hspace{3.7mm}}c@{\hspace{3.7mm}}c@{\hspace{3.7mm}}c@{\hspace{3.7mm}}c@{\hspace{3.7mm}}c@{\hspace{3.7mm}}c@{\hspace{3.7mm}}c@{\hspace{3.7mm}}c@{\hspace{3.7mm}}c@{\hspace{3.7mm}}c@{\hspace{3.7mm}}c@{\hspace{3.7mm}}c@{\hspace{3.7mm}}c@{\hspace{3.7mm}}c}
Star &1&2&3&4&5&6&7&8&9&10&11&12&13&14&15&16&17&18&19&20&21&22&23&24\\
\hline
V1&+&-&+&+&-&+&+&-&-&-&+&-&+&-&+&+&+&-&{*}&+&+&+&-&+\\
....&....&....&....&....&....&....&....&....&....&....&....&....&....&....&....&....&....&....&....&....&....&....&....&....\\
\hline
\end{tabular}
\end{minipage}
\end{table*}

\subsection{The method}
\label{method}
The method used for the analysis of period changes is the same as employed by \citet{ocen} and \citet{m5} to study the period changes of RR Lyrae stars in $\omega$ Cen and M5, respectively.

The original light-curve data (time series) were used for the analysis. Photographic, CCD $B$ ($g$), and CCD $V$ data if no simultaneous CCD $B$ observations had been obtained, were utilized. The datasets of the variables used for the period change study are listed in Table \ref{felhasznalt}. For each star, the dataset that defines the mean (master) light curve is marked by an asterisk. The complete table is available as supporting information in the electronic version of this article.

The mean light curves were constructed in the form of fifth and seventh order Fourier series for the RRc and RRab stars, respectively. In case of noisy and/or strongly variable light curves, lower order fits were applied to avoid unreal, wavy mean light-curve shapes.

As the first step, each dataset was magnitude homogenized by fitting the mean light curve. In the second step, the mean light curve was fitted in phase to 2--4 years long segments of the magnitude homogenized, combined light curve. By default, the data were divided into 19 segments, and their phase-shift values corresponded to the $O-C$ for the mean epoch of the data subsets. Some of the bins were divided into shorter segments in case of very fast period changes, or they were merged together if insufficient number of data points were available in one of the bins, and no significant period change occurred in the given time interval.

The $O-C$ values, derived in this way, were used to construct the phase-shift diagrams ($O-C$ diagrams). Since there were gaps in the coverage of the observations, cycle count ambiguities occurred in some cases, especially for stars showing irregular, large period changes. These $O-C$ points were adjusted upward or downward by integer cycle numbers to obtain the most acceptable period-change solution (see details in the next section). Then, the magnitude homogenization was refined as described in section 2.3 of \citet{m5}. For the stars with abrupt period changes, this step greatly improved the magnitude homogenization. The determination of the $O-C$ points and the construction of the $O-C$ diagrams were finalized on the refined version of the magnitude-homogenized data.

The $O-C$ diagrams were fitted by polynomials, and the period-change rates of the variables were measured by the coefficients of these functions. The reliability of the $O-C$ fits were checked by the comparison of the results of direct period determinations with the derivative of the $O-C$ fits, and by the construction of folded light curves corrected for the period change.

Table \ref{ocvalues} gives an example of the results for V1. The time interval, the corresponding  mean epoch, the number of datapoints of the subsets and the  $O-C$ or temporal period values and their errors are given in the columns.  The complete table, including the $O-C$ values and instantaneous periods for all the analyzed stars is available in the electronic version of the article as supporting information.

\begin{table*}
\caption{$O-C$ values and temporal periods for V1. The complete table including all the variables is given in the electronic version of the article as supporting information.}
 \label{ocvalues}
  \begin{tabular}{c@{\hspace{6mm}}cr@{\hspace{6mm}}r@{\hspace{6mm}}c}
  \hline
\multicolumn{1}{l}{V1}& \multicolumn{2}{c}{  $P_a$=0.52061350 d}\\
 \hline
time interval [JD$-2400000$]&  $\overline{JD}$ &$N$   &$O-C$ [d]	&   error [1$\sigma$]\\
\hline
13372.5--14079.7& 13804.0& 16&	$-0.2475$& 0.0128\\
14437.5--15161.9& 15044.7& 32&	$-0.1843$& 0.0059\\
22455.4--22840.6& 22750.7& 131&	0.0764& 0.0011\\
23858.5--24684.9& 24470.8& 157&	0.1268& 0.0011\\
28963.4--30078.6& 29736.9& 31&	0.2083& 0.0029\\
33034.5--33763.6& 33486.2& 44&	0.2356& 0.0020\\
34076.5--34834.5& 34317.8& 94&	0.2423& 0.0010\\
35219.3--35933.6& 35559.0& 133&	0.2478& 0.0015\\
36367.3--37131.4& 36772.7& 95&	0.2446& 0.0017\\
37757.5--38502.6& 38060.7& 21&	0.2262& 0.0035\\
42838.5--43254.6& 43042.2& 60&	0.1603& 0.0018\\
43597.3--44343.5& 43912.7& 50&	0.1812& 0.0059\\
47276.3--47555.6& 47505.2& 51&	0.0450& 0.0022\\
48755.6--49092.0& 48939.0& 183&	0.0001& 0.0005\\
50162.4--50176.8& 50170.9& 169&	$-0.0403$& 0.0010\\
50896.5--51283.6& 50964.6& 233&	$-0.0670$& 0.0003\\
54869.5--54965.5& 54929.0& 725&	$-0.2271$& 0.0005\\
\hline
time interval [JD$-2400000$]&     $\overline{JD}$  & $N$&     $P$ [d]      & error [1$\sigma$]\\
\hline
13372.5--15161.9& 14631.1& 48&	0.5206534& 0.0000039\\
22455.4--24684.9& 23688.4& 288&	0.5206284& 0.0000005\\
28963.4--34567.4& 33232.6& 166&	0.5206175& 0.0000003\\
34826.3--38502.6& 36216.4& 252&	0.5206096& 0.0000006\\
42838.5--44343.5& 43437.9& 110&	0.5206131& 0.0000032\\
47276.3--49092.0& 48626.5& 234&	0.5205958& 0.0000005\\
50162.4--54965.5& 53395.9& 1127& 0.5205926& 0.0000001\\
\hline
\end{tabular}
\end{table*}

\section{$O-C$ diagrams}
\label{ocsect}
The results of the period-change analysis are documented in Fig.~\ref{oc} for 129 RR Lyrae stars. The variables are shown in the order of the length of their periods. Because of their extremely strong phase modulations (Blazhko effect) and abrupt changes of the pulsation period, no period-change solution could be obtained for three RRab variables (V5, V50 and V130). The strong changes of their modal content made the analysis of V79 and V99 impossible using our method. Thus,  $O-C$ diagrams have not been constructed for five of the studied 134 variables.

Three panels are shown for each variable. The left-hand panels show the $O-C$ diagrams and their polynomial fits. 
Variables showing Blazhko effect or double-mode pulsations are denoted as `Bl' and `dm', respectively.

The differences between temporal period values ($P$) determined by direct period analysis of the data subsets and the mean period ($P_a$) adopted to calculate $O-C$ values are plotted in the middle panels [$(P-P_a)\times 10^5$ d]. The variations of the instantaneous periods [$P(t)$] predicted from the $O-C$ data using Eq. 3 of \citet{ocen} are also shown for comparison. When cycle-count ambiguity of the $O-C$ values occurred, that solution was accepted, which yielded the smallest difference of the directly observed period values from the $P(t)$ function.

The errorbars shown for the $O-C$ points and the observed period values indicate $2\sigma$ formal uncertainties determined by least-squares fitting method. 

The right-hand panels show the folded light curves of the time-transformed (Eq. 4 of \citet{ocen}) data, which are corrected for the period variations according to the polynomial fits of the $O-C$ diagrams.

The $O-C$ solutions are shown in two separate plots using different periods for V13 and V41, as the period changes of these stars are too large to be shown in a single plot.

The very irregular $O-C$ variations of thirteen stars (V12, V17, V18, V28, V34, V35, V43, V44, V47, V54, V70, V80 and V110) could be fitted only by two consecutive, separate polynomials.

\begin{figure*}
\includegraphics[width=17.5cm]{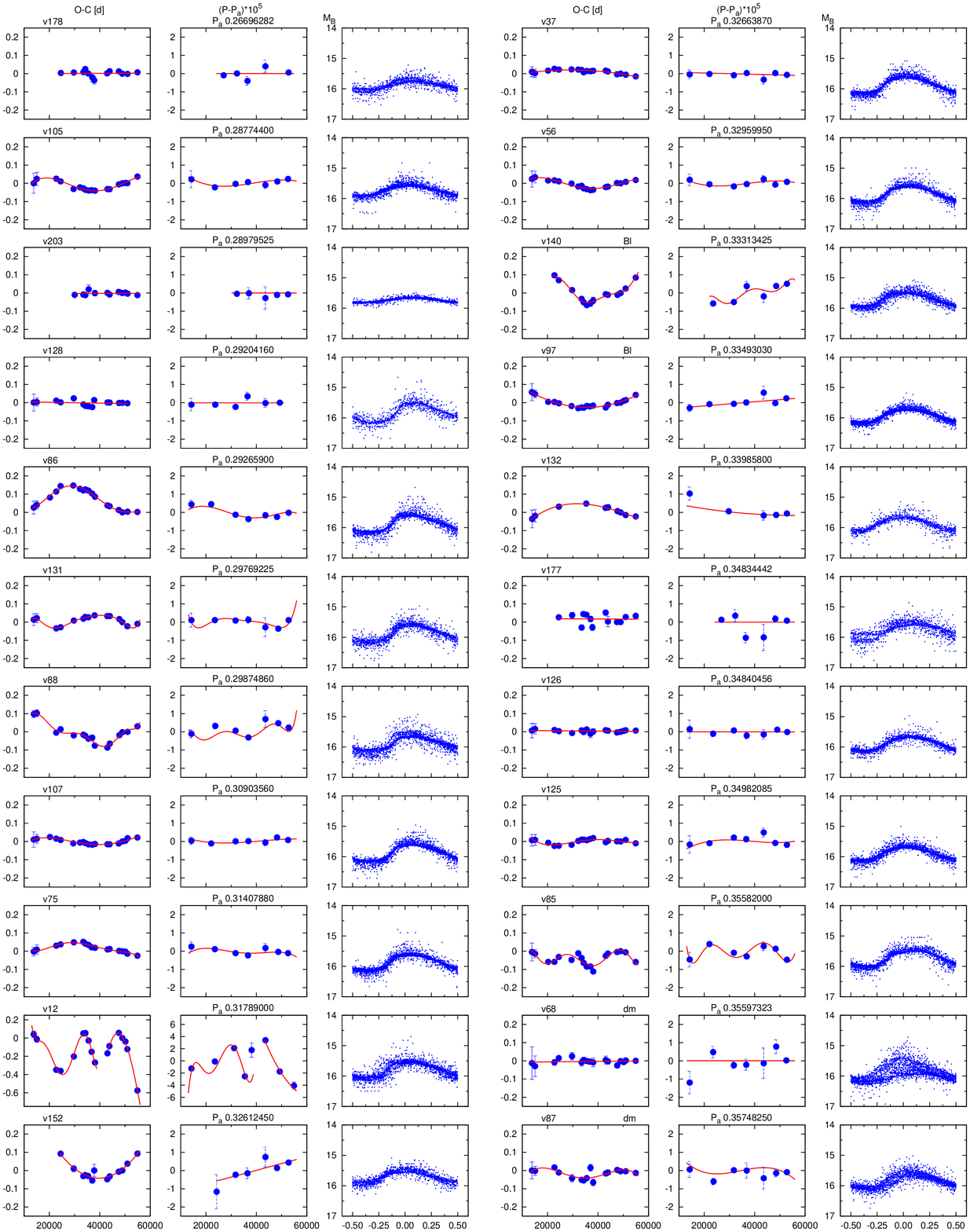}
\caption{$O-C$ diagrams, period changes and phased light curves of the data corrected for the fitted $O-C$ variations for 129 RR Lyrae stars in M3. See Sect.~\ref{ocsect} for details.}
\label{oc}
\end{figure*}

\begin{figure*}
\includegraphics[width=17.7cm]{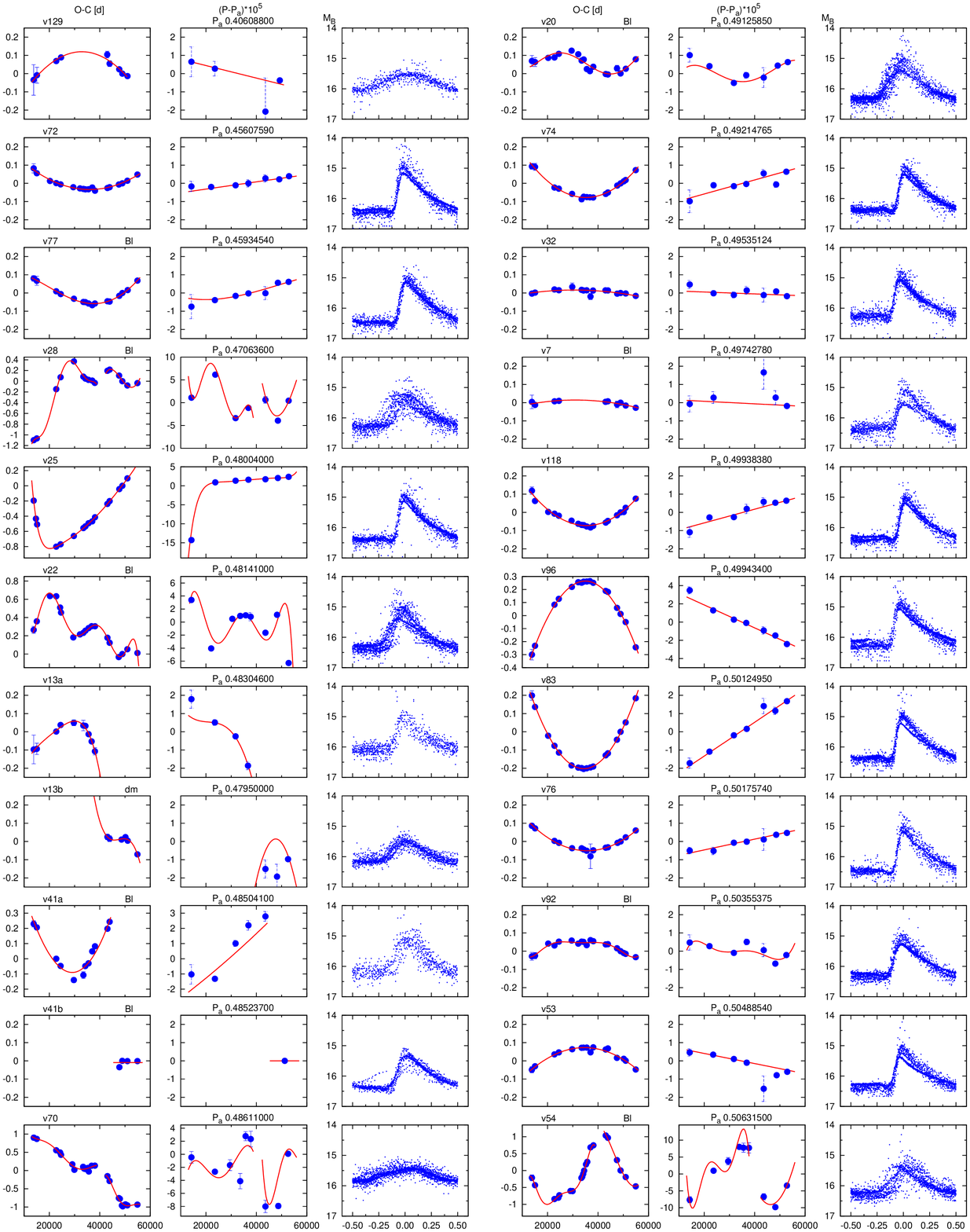}
\contcaption{}
\end{figure*}
\begin{figure*}
\includegraphics[width=17.7cm]{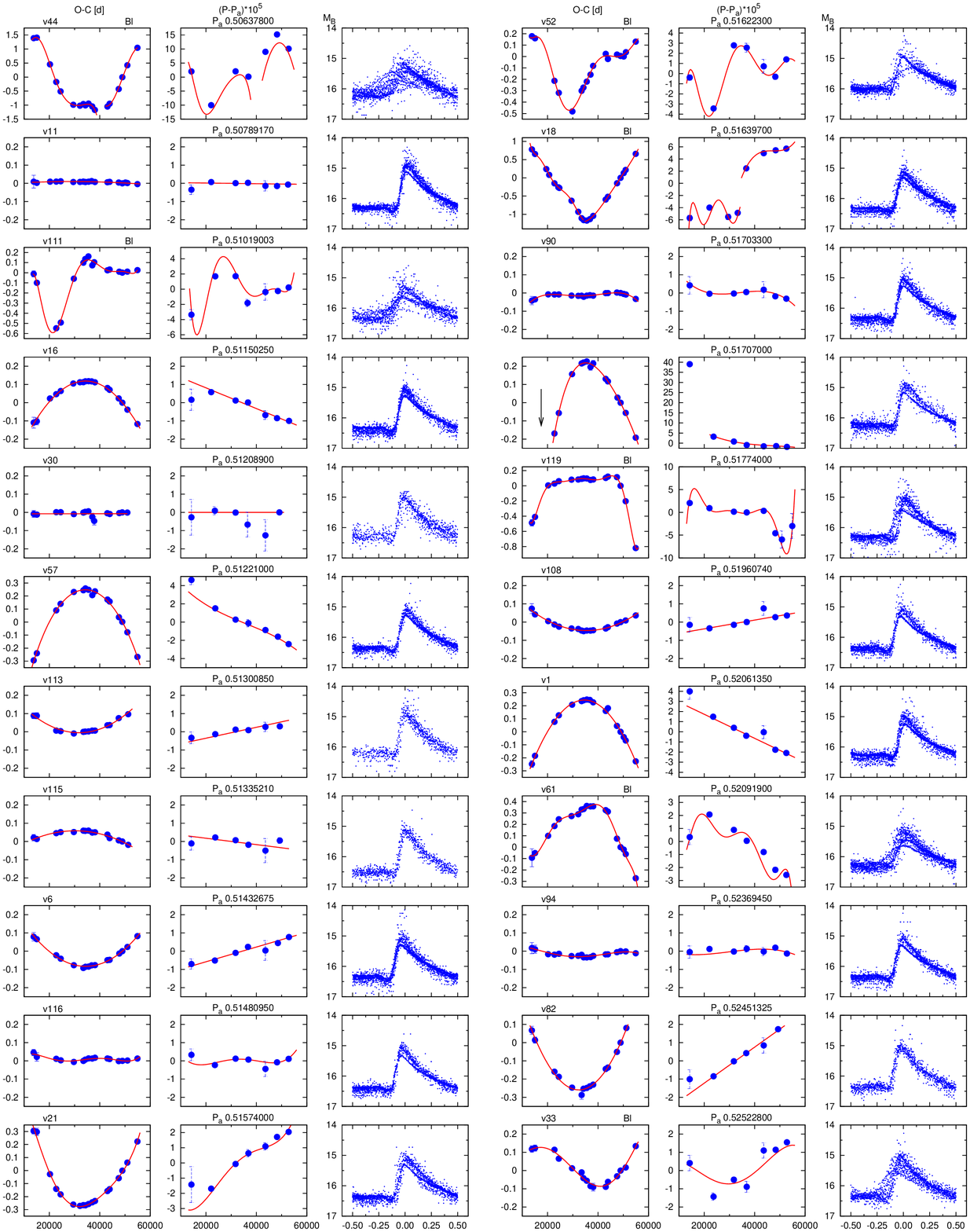}
\contcaption{}
\end{figure*}
\begin{figure*}
\includegraphics[width=17.7cm]{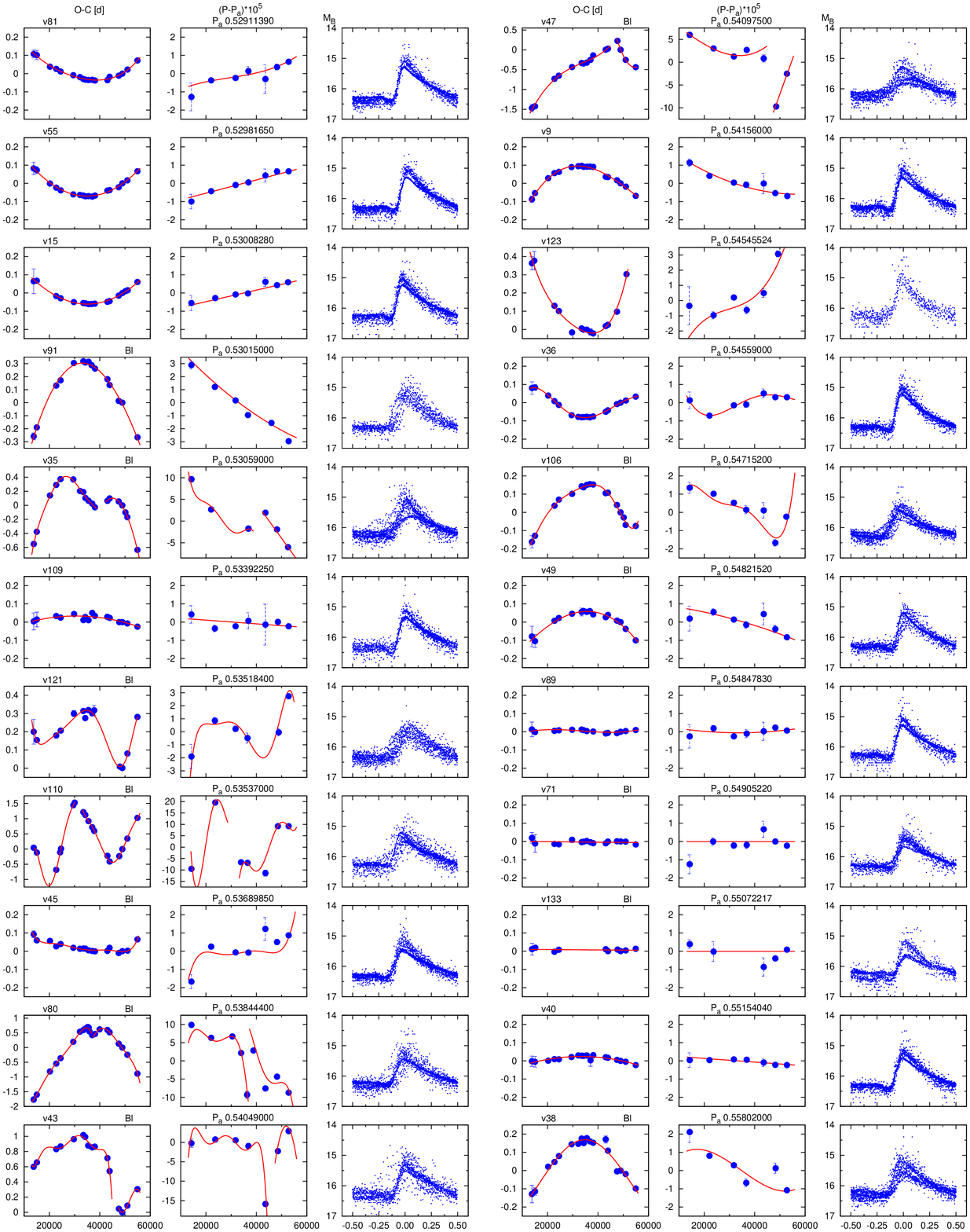}
\contcaption{}
\end{figure*}
\begin{figure*}
\includegraphics[width=17.7cm]{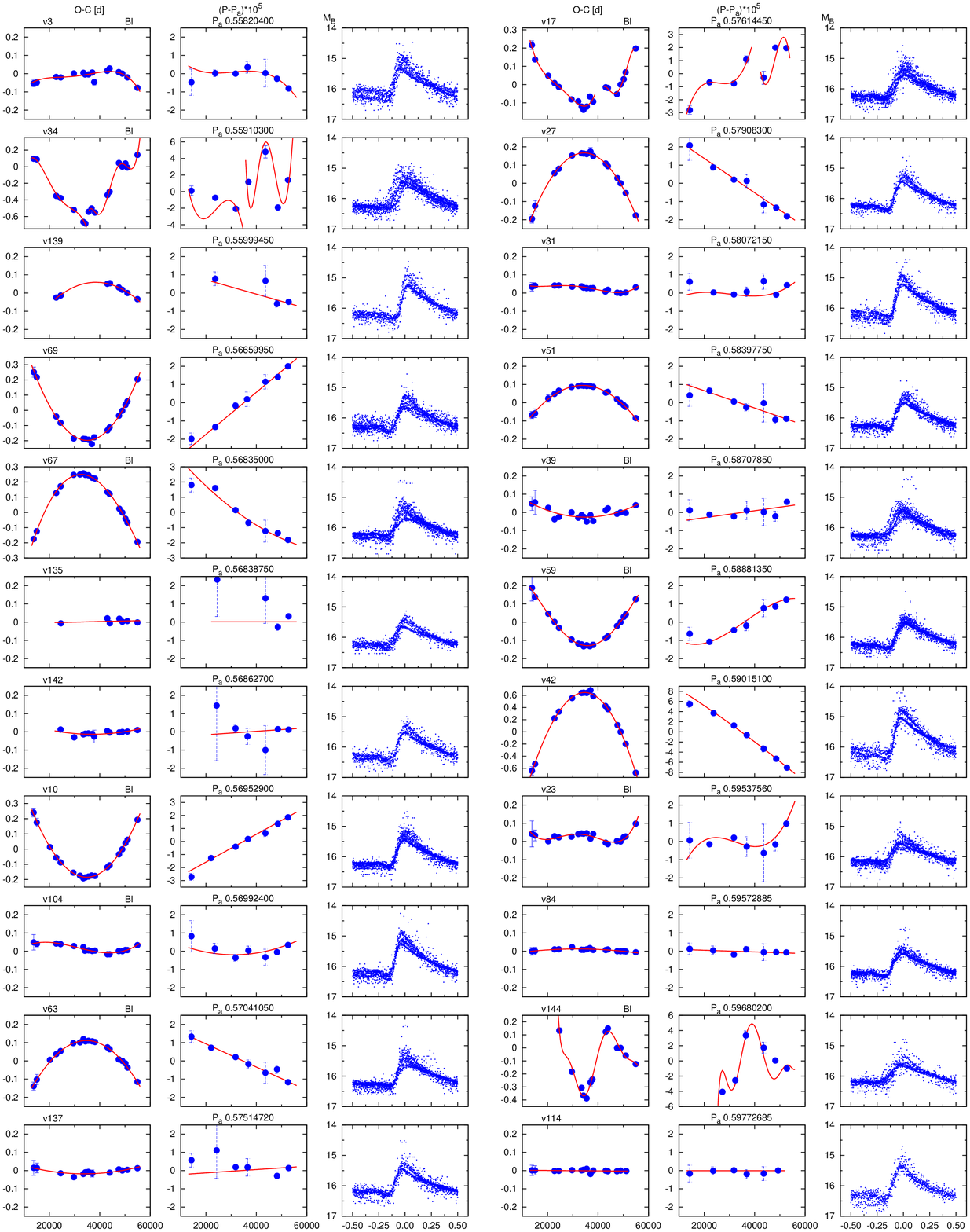}
\contcaption{}
\end{figure*}
\begin{figure*}
\includegraphics[width=17.7cm]{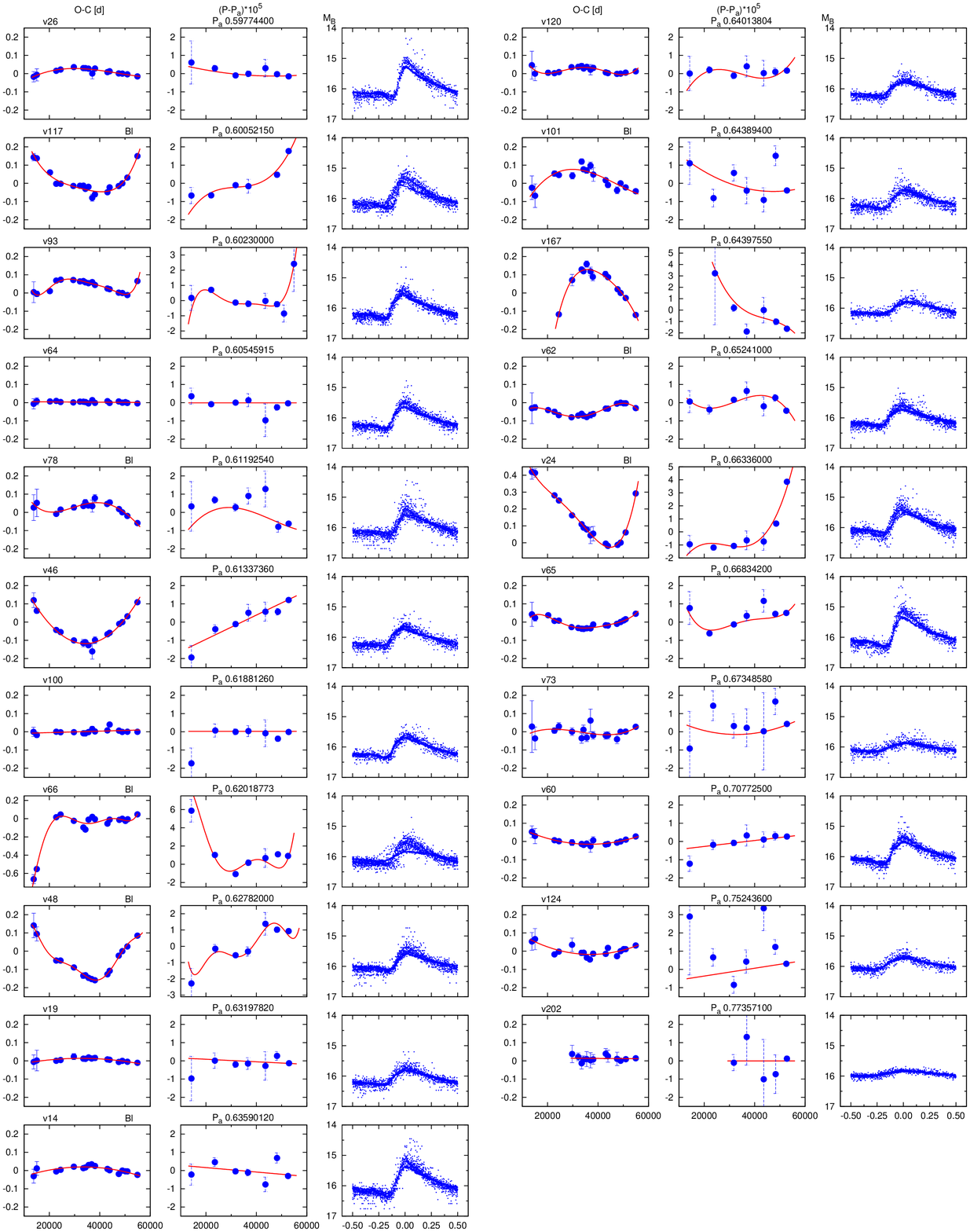}
\contcaption{}
\end{figure*}

\section{Results and Discussion}
\subsection{Period-change results}
\label{tablasect}
\begin{table*}
\caption{Summary of the light-curve parameters and period-change properties of 134 RR Lyrae stars in M3.}
\begin{minipage}{\textwidth}
\label{tabla}
\begin{tabular}{l@{\hspace{2mm}}ccccccrrrrr}
Star$^{a}$&	&	$P_a^{b,c}$		& $V_\mathrm{i}^d$		&	$A_V^{e,f}$	&Blazhko&Type$^{g}$&	\multicolumn{2}{c}{$\beta=\dot{P}$} & \multicolumn{2}{c}{$\alpha=P^{-1}\dot{P}$}  & $(P_{max}-P_{min})P_{a}^{-1}$\\
\hline
	&			&	d 			&\multicolumn{2}{c}{mag}&		&		& $10^{-10}$& $\mathrm{d Myr^{-1}}$& $\mathrm{10^{-10}d^{-1}}$& $\mathrm{Myr^{-1}}$	&$10^{-5}$\\
\hline
V1	&	RRab	&	0.52061350	&	15.632	&	1.17	&	--	&	n		&	-11.42	&-0.417	&-21.93	&-0.801	&11.68 (0.75)\\	
V3	&	RRab	&	0.55820400	&	15.590	&	1.17	&	+	&	i		&	--		&--		&--		&	--&2.20 (0.31)\\	
V5$^*$	&	RRab	&	0.50570380	&	15.635	&	0.80	&	+	&	ii		&	--		&--		&--			&--&	--\\	
V6	&	RRab	&	0.51432675	&	15.705	&	1.13	&	--	&	p		&	3.93	&0.144	&7.64	&0.279		&2.88 (0.27)\\	
V7	&	RRab	&	0.49742780	&	15.648	&	1.12	&	+	&	n		&	-0.70	&-0.025	&-1.40	&-0.051		&3.70 (0.93)\\	
V9	&	RRab	&	0.54156000	&	15.647	&	1.06	&	--	&	$\textrm{n}_{\textrm{i}}$	&	\textit{-4.00}	&\textit{-0.146}	&\textit{-7.39}	&\textit{-0.270}	&	3.38 (0.19)\\	
V10	&	RRab	&	0.56952900	&	15.654	&	0.89	&	+	&	p		&	9.39	&0.343	&16.48	&0.602		&8.04 (0.23)\\	
V11	&	RRab	&	0.50789170	&	15.631	&	1.26	&	--	&	n		&	-0.18	&-0.006	&-0.35	&-0.013		&0.83 (0.26)\\	
V12	&	RRc		&	0.31789000	&	15.606	&	0.49	&	--	&	i		&	--		&--		&--		&	--		&23.56 (1.13)\\	
V13	&	RRd		&	0.47950000	&	15.699	&	0.68	&	--	&	i		&	--		&--		&--		&	--		&746.76 (0.86)\\	
V14	&	RRab	&	0.63590120	&	15.539	&	0.90	&	+	&	n		&	-0.96	&-0.035	&-1.51	&-0.055		&2.28 (0.38)\\	
V15	&	RRab	&	0.53008280	&	15.617	&	1.09	&	--	&	p		&	3.00	&0.109	&5.65	&0.206		&2.19 (0.46)\\	
V16	&	RRab	&	0.51150250	&	15.704	&	1.17	&	--	&	n		&	-5.56	&-0.203	&-10.87	&-0.397		&3.09 (0.06)\\	
V17	&	RRab	&	0.57614450	&	15.651	&	0.80	&	+	&	i		&	--		&--		&--		&	--		&8.31 (0.30)\\	
V18	&	RRab	&	0.51639700	&	15.701	&	1.14	&	+	&	i		&	--		&--		&--		&	--		&22.11 (0.45)\\	
V19	&	RRab	&	0.63197820	&	15.685	&	0.45	&	--	&	n		&	-0.55	&-0.020	&-0.86	&-0.032		&1.96 (0.98)\\	
V20	&	RRab	&	0.49125850	&	15.635	&	0.93	&	+	&	i		&	--		&--		&--		&	--		&4.58 (0.45)\\	
V21	&	RRab	&	0.51574000	&	15.671	&	1.16	&	--	&	$\textrm{p}_{\textrm{i}}$	&	\textit{12.98}	&\textit{0.474}	&\textit{25.16}	&\textit{0.919}	&	8.28 (0.91)\\	
V22	&	RRab	&	0.48141000	&	15.715	&	1.03	&	+	&	i		&	--		&--		&--		&	--		&20.11 (0.24)\\	
V23	&	RRab	&	0.59537560	&	15.629	&	0.70	&	+	&	i		&	--		&--		&--		&	--		&2.70 (1.33)\\	
V24	&	RRab	&	0.66336000	&	15.531	&	0.69	&	+	&	i		&	--		&--		&--		&	--		&7.61 (0.15)\\	
V25	&	RRab	&	0.48005640	&	15.675	&	1.20	&	--	&	i		&	--		&--		&--		&	--		&34.68 (0.50)\\	
V26	&	RRab	&	0.59774400	&	15.602	&	0.99	&	--	&	$\textrm{n}_{\textrm{i}}$	&	\textit{-0.98}	&\textit{-0.036}	&\textit{-1.65}	&\textit{-0.060}	&	1.27 (0.99)\\	
V27	&	RRab	&	0.57908300	&	15.653	&	0.88	&	--	&	n		&	-8.14	&-0.297	&-14.05	&-0.513		&6.72 (0.71)\\	
V28	&	RRab	&	0.47060900	&	15.717	&	0.82	&	+	&	i		&	--		&--		&--		&	--		&21.42 (0.42)\\	
V30	&	RRab	&	0.51208900	&	15.675	&	1.13	&	--	&	s		&	0.00	&0.000	&0.00	&0.000		&2.64 (0.87)\\	
V31	&	RRab	&	0.58072150	&	15.554	&	1.18	&	--	&	$\textrm{s}_{\textrm{i}}$	&	\textit{0.00}	&\textit{0.000}	&\textit{0.00}	&\textit{0.000}	&	1.27 (0.39)\\	
V32	&	RRab	&	0.49535124	&	15.672	&	1.17	&	--	&	n		&	-0.53	&-0.019	&-1.07	&-0.039		&1.29 (0.22)\\	
V33	&	RRab	&	0.52522800	&	15.639	&	0.92	&	+	&	i		&	--		&--		&--		&	--		&5.67 (0.19)\\	
V34	&	RRab	&	0.55910310	&	15.666	&	0.74	&	+	&	i		&	--		&--		&--		&	--		&12.31 (0.68)\\	
V35	&	RRab	&	0.53059000	&	15.656	&	0.72	&	+	&	i		&	--		&--		&--		&	--		&29.55 (0.57)\\	
V36	&	RRab	&	0.54559000	&	15.606	&	1.20	&	--	&	i		&	--		&--		&--		&	--		&2.24 (0.22)\\	
V37	&	RRc		&	0.32663870	&	15.682	&	0.49	&	--	&	n		&	-0.54	&-0.020	&-1.66	&-0.061		&1.10 (0.39)\\	
V38	&	RRab	&	0.55802000	&	15.692	&	0.80	&	+	&	$\textrm{n}_{\textrm{i}}$	&	\textit{-6.78}	&\textit{-0.248}	&\textit{-12.15}	&\textit{-0.444}	&	5.70 (0.50)\\	
V39	&	RRab	&	0.58707850	&	15.684	&	0.75	&	+	&	p		&	1.64	&0.060	&2.80	&0.102		&1.36 (0.09)\\	
V40	&	RRab	&	0.55154040	&	15.677	&	0.98	&	--	&	n		&	-0.92	&-0.034	&-1.67	&-0.061		&0.56 (0.04)\\	
V41	&	RRab	&	0.48523700	&	15.654	&	0.88	&	+	&	i		&	--		&--		&--		&	--		&43.11 (0.15)\\	
V42	&	RRab	&	0.59015100	&	15.495	&	1.25	&	--	&	$\textrm{n}_{\textrm{i}}$	&	\textit{-30.99}	&\textit{-1.132}	&\textit{-52.52}	&\textit{-1.918}	&	21.27 (0.54)\\	
V43	&	RRab	&	0.54049000	&	15.678	&	1.00	&	+	&	i		&	--		&--		&--		&	--		&34.78 (0.63)\\	
V44	&	RRab	&	0.50640000	&	15.624	&	0.70	&	+	&	i		&	--		&--		&--		&	--		&49.72 (0.74)\\	
V45	&	RRab	&	0.53689850	&	15.698	&	0.90	&	+	&	i		&	--		&--		&--		&	--		&5.36 (0.68)\\	
V46	&	RRab	&	0.61337360	&	15.703	&	0.54	&	--	&	p		&	5.42	&0.198	&8.84	&0.323		&5.14 (0.54)\\	
V47	&	RRab	&	0.54097500	&	15.693	&	0.65	&	+	&	i		&	--		&--		&--		&	--		&29.02 (0.62)\\	
V48	&	RRab	&	0.62782000	&	15.613	&	0.57	&	+	&	i		&	--		&--		&--		&	--		&5.83 (0.92)\\	
V49	&	RRab	&	0.54821520	&	15.647	&	1.10	&	+	&	$\textrm{n}_{\textrm{i}}$	&	\textit{-3.60}	&\textit{-0.132}	&\textit{-6.57}	&\textit{-0.240}	&	2.52 (0.17)\\	
V50$^*$	&	RRab	&	0.51289670	&	15.679	&	0.75	&	+	&	ii		&	--		&--		&--		&--		&	--\\	
V51	&	RRab	&	0.58397750	&	15.662	&	0.85	&	--	&	n		&	-4.08	&-0.149	&-6.98	&-0.255		&2.72 (0.23)\\	
V52	&	RRab	&	0.51622300	&	15.666	&	0.90	&	+	&	i		&	--		&--		&--		&--			&12.07 (0.17)\\	
V53	&	RRab	&	0.50488540	&	15.671	&	1.18	&	--	&	n		&	-2.75	&-0.100	&-5.44	&-0.199		&3.94 (0.72)\\	
V54	&	RRab	&	0.50631500	&	15.683	&	0.79	&	+	&	i		&	--		&--		&--		&--			&35.18 (0.55)\\	
V55	&	RRab	&	0.52981650	&	15.677	&	1.09	&	--	&	p		&	3.44	&0.125	&6.48	&0.237		&3.13 (0.38)\\	
V56	&	RRc		&	0.32959950	&	15.636	&	0.47	&	--	&	i		&	--		&--		&--		&	--		&1.21 (0.31)\\	
V57	&	RRab	&	0.51221000	&	15.694	&	1.10	&	--	&	$\textrm{n}_{\textrm{i}}$	&	\textit{-12.57}	&\textit{-0.459}	&\textit{-24.53}	&\textit{-0.896}	&	13.72 (0.51)\\	
V58	&	RRab	&	0.51707000	&	15.597	&	1.29	&	--	&	i		&	--		&--		&--		&	--		&78.91 (0.19)\\	
V59	&	RRab	&	0.58881350	&	15.673	&	0.82	&	+	&	$\textrm{p}_{\textrm{i}}$	&	\textit{6.68}	&\textit{0.244}	&\textit{11.35}	&\textit{0.415}	&	3.92 (0.09)\\	
V60	&	RRab	&	0.70772500	&	15.505	&	0.69	&	--	&	p		&	1.16	&0.042	&1.63	&0.060		&2.19 (0.51)\\	
V61	&	RRab	&	0.52091900	&	15.703	&	0.82	&	+	&	i		&	--		&--		&--		&	--		&8.91 (0.10)\\	
V62	&	RRab	&	0.65241000	&	15.639	&	0.52	&	+	&	i		&	--		&--		&--		&	--		&1.66 (0.38)\\	
V63	&	RRab	&	0.57041050	&	15.688	&	0.74	&	+	&	n		&	-5.57	&-0.203	&-9.76	&-0.356		&4.38 (0.28)\\	
V64	&	RRab	&	0.60545915	&	15.670	&	0.71	&	--	&	s		&	0.00	&0.000	&0.00	&0.000		&2.18 (0.83)\\		
\end{tabular}
\end{minipage}
\end{table*}
\begin{table*}
\begin{minipage}{\textwidth}
\contcaption{}
\begin{tabular}{l@{\hspace{2mm}}ccccccrrrrr}
Star$^{a}$&	&	$P_a^{b,c}$		& $V_\mathrm{i}^d$		&	$A_V^{e,f}$	&Blazhko&Type$^{g}$&	\multicolumn{2}{c}{$\beta=\dot{P}$} & \multicolumn{2}{c}{$\alpha=P^{-1}\dot{P}$}  & $(P_{max}-P_{min})P_{a}^{-1}$\\
\hline
	&			&	d 			&\multicolumn{2}{c}{mag}&		&		& $10^{-10}$& $\mathrm{d Myr^{-1}}$& $\mathrm{10^{-10}d^{-1}}$& $\mathrm{Myr^{-1}}$	&$10^{-5}$\\
\hline
V65	&	RRab	&	0.66834200	&	15.502	&	0.92	&	--	&	i		&	--		&--		&--		&	--		&2.66 (0.47)\\	
V66	&	RRab	&	0.62018773	&	15.646	&	0.50	&	+	&	i		&	--		&--		&--		&	--		&11.21 (1.01)\\	
V67	&	RRab	&	0.56835000	&	15.673	&	0.75	&	+	&	$\textrm{n}_{\textrm{i}}$	&	\textit{-10.41}	&\textit{-0.380}	&\textit{-18.32}	&\textit{-0.669}	&	6.35 (0.41)\\	
V68	&	RRd		&	0.35597323	&	15.626	&	0.39	&	--	&	s		&	0.00	&0.000	&0.00	&0.000		&5.53 (1.02)\\	
V69	&	RRab	&	0.56659950	&	15.677	&	0.89	&	--	&	p		&	10.20	&0.373	&18.01	&0.658		&7.02 (0.28)\\	
V70	&	RRc		&	0.48611000	&	15.391	&	0.35	&	--	&	i		&	--		&--		&--		&	--		&22.20 (1.22)\\	
V71	&	RRab	&	0.54905220	&	15.687	&	0.81	&	+	&	s		&	0.00	&0.000	&0.00	&0.000		&3.50 (0.62)\\	
V72	&	RRab	&	0.45607590	&	15.670	&	1.30	&	--	&	p		&	2.18	&0.080	&4.78	&0.175		&1.29 (0.07)\\	
V73	&	RRab	&	0.67348580	&	15.624	&	0.25	&	+	&	i		&	--		&--		&--		&	--		&3.83 (1.61)\\	
V74	&	RRab	&	0.49214765	&	15.685	&	1.26	&	--	&	p		&	3.91	&0.143	&7.95	&0.290		&3.29 (0.63)\\	
V75	&	RRc		&	0.31407880	&	15.631	&	0.49	&	--	&	i		&	--		&--		&--		&	--		&1.53 (0.46)\\	
V76	&	RRab	&	0.50175740	&	15.716	&	1.15	&	--	&	p		&	2.95	&0.108	&5.88	&0.215		&1.97 (0.20)\\	
V77	&	RRab	&	0.45934540	&	15.722	&	1.31	&	+	&	$\textrm{p}_{\textrm{i}}$	&	\textit{2.99}	&\textit{0.109}	&\textit{6.52}	&\textit{0.238}	&	2.96 (0.72)\\	
V78	&	RRab	&	0.61192540	&	15.572	&	0.73	&	+	&	i		&	--		&--		&--		&	--		&3.38 (0.84)\\	
V79$^{*}$	&	RRab	&	0.48330344	&	15.721	&	1.18	&	+	&	ii		&	--		&--		&--		&--		&	--\\	
V80	&	RRab	&	0.53837580	&	15.639	&	0.85	&	+	&	i		&	--		&--		&--		&	--		&35.57 (1.06)\\	
V81	&	RRab	&	0.52911390	&	15.678	&	1.08	&	--	&	$\textrm{p}_{\textrm{i}}$	&	\textit{2.81}	&\textit{0.103}	&\textit{5.32}	&\textit{0.194}	&	3.65 (0.72)\\	
V82	&	RRab	&	0.52451310	&	15.649	&	1.12	&	--	&	p		&	9.43	&0.344	&17.97	&0.656		&5.22 (0.50)\\	
V83	&	RRab	&	0.50124950	&	15.679	&	1.20	&	--	&	p		&	9.25	&0.338	&18.46	&0.674		&6.78 (0.28)\\	
V84	&	RRab	&	0.59572885	&	15.660	&	0.73	&	--	&	n		&	-0.39	&-0.014	&-0.65	&-0.024		&0.52 (0.29)\\	
V85	&	RRc		&	0.35582000	&	15.575	&	0.51	&	--	&	i		&	--		&--		&--		&	--		&2.42 (0.06)\\	
V86	&	RRc		&	0.29265900	&	15.668	&	0.53	&	--	&	i		&	--		&--		&--		&	--		&2.80 (0.15)\\	
V87	&	RRd		&	0.35748250	&	15.577	&	0.43	&	--	&	i		&	--		&--		&--		&	--		&1.34 (0.26)\\	
V88	&	RRc		&	0.29874860	&	15.685	&	0.57	&	--	&	i		&	--		&--		&--		&	--		&3.38 (0.80)\\	
V89	&	RRab	&	0.54847830	&	15.644	&	1.00	&	--	&	$\textrm{s}_{\textrm{i}}$	&	\textit{0.00}	&\textit{0.000}	&\textit{0.00}	&\textit{0.000}	&	0.88 (0.11)\\	
V90	&	RRab	&	0.51703300	&	15.676	&	1.15	&	--	&	$\textrm{s}_{\textrm{i}}$	&	\textit{0.00}	&\textit{0.000}	&\textit{0.00}	&\textit{0.000}	&	1.41 (0.46)\\	
V91	&	RRab	&	0.53015000	&	15.637	&	1.08	&	+	&	$\textrm{n}_{\textrm{i}}$	&	\textit{-13.58}	&\textit{-0.496}	&\textit{-25.61}	&\textit{-0.936}	&	11.03 (0.30)\\	
V92	&	RRab	&	0.50355375	&	15.639	&	1.02	&	+	&	i		&	--		&--		&--		&	--		&2.36 (0.20)\\	
V93	&	RRab	&	0.60230000	&	15.637	&	0.74	&	--	&	i		&	--		&--		&--		&	--		&5.43 (1.58)\\	
V94	&	RRab	&	0.52369450	&	15.678	&	1.12	&	--	&	$\textrm{s}_{\textrm{i}}$	&	\textit{0.00}	&\textit{0.000}	&\textit{0.00}	&\textit{0.000}	&	0.61 (0.08)\\	
V96	&	RRab	&	0.49943400	&	15.653	&	1.24	&	--	&	n		&	-12.59	&-0.460	&-25.20	&-0.921		&11.79 (0.38)\\	
V97	&	RRc		&	0.33493030	&	15.707	&	0.40	&	+	&	p		&	1.71	&0.063	&5.11	&0.187		&2.51 (0.61)\\	
V99$^*$	&	RRd		&	0.36083550	&	15.596	&	0.37	&	--	&	ii		&	--		&--		&--		&--		&	--\\	
V100	&	RRab	&	0.61881260	&	15.704	&	0.64	&	--	&	s		&	0.00	&0.000	&0.00	&0.000		&2.91 (0.74)\\	
V101	&	RRab	&	0.64389400	&	15.673	&	0.54	&	+	&	$\textrm{n}_{\textrm{i}}$	&	\textit{-3.02}	&\textit{-0.110}	&\textit{-4.69}	&\textit{-0.171}	&	3.76 (0.66)\\	
V104	&	RRab	&	0.56992400	&	15.554	&	1.11	&	+	&	i		&	--		&--		&--		&	--		&2.07 (0.77)\\	
V105	&	RRc		&	0.28774400	&	15.583	&	0.30	&	--	&	i		&	--		&--		&--		&	--		&1.60 (0.14)\\	
V106	&	RRab	&	0.54714500	&	15.645	&	0.73	&	+	&	i		&	--		&--		&--		&	--		&5.52 (0.32)\\	
V107	&	RRc		&	0.30903560	&	15.677	&	0.52	&	--	&	i		&	--		&--		&--		&	--		&1.07 (0.12)\\	
V108	&	RRab	&	0.51960740	&	15.700	&	1.14	&	--	&	p		&	2.32	&0.085	&4.46	&0.163		&2.10 (0.37)\\	
V109	&	RRab	&	0.53392250	&	15.693	&	1.12	&	--	&	n		&	-0.96	&-0.035	&-1.79	&-0.065		&1.44 (0.48)\\	
V110	&	RRab	&	0.53537000	&	15.699	&	1.00	&	+	&	i		&	--		&--		&--		&	--		&57.82 (1.26)\\	
V111	&	RRab	&	0.51019003	&	15.687	&	0.91	&	+	&	i		&	--		&--		&--		&	--		&9.94 (0.71)\\	
V113	&	RRab	&	0.51300850	&	--	&	\textit{1.10}&	--	&	p		&	2.92	&0.107	&5.70	&0.208		&1.23 (0.31)\\	
V114	&	RRab	&	0.59772685	&	15.689	&	0.84	&	--	&	s		&	0.00	&0.000	&0.00	&0.000		&0.35 (0.22)\\	
V115	&	RRab	&	0.51335210	&	--	&	\textit{1.15}&	--	&	n		&	-1.70	&-0.062	&-3.32	&-0.121		&1.40 (0.65)\\	
V116	&	RRab	&	0.51480950	&	15.666	&	1.15	&	--	&	$\textrm{s}_{\textrm{i}}$	&	\textit{0.00}	&\textit{0.000}	&\textit{0.00}	&\textit{0.000}	&	1.34 (0.54)\\	
V117	&	RRab	&	0.60052150	&	15.632	&	0.83	&	+	&	$\textrm{p}_{\textrm{i}}$	&	\textit{4.12}	&\textit{0.150}	&\textit{6.86}	&\textit{0.250}	&	4.06 (0.37)\\	
V118	&	RRab	&	0.49938380	&	15.692	&	1.21	&	--	&	p		&	3.77	&0.138	&7.54	&0.276		&3.46 (0.28)\\	
V119$^{*}$	&	RRab	&	0.51774000	&	15.635	&	0.98	&	+	&	i		&	--		&--		&--		&	--		&15.47 (1.83)\\	
V120	&	RRab	&	0.64013804	&	15.625	&	0.42	&	--	&	$\textrm{s}_{\textrm{i}}$	&	\textit{0.00}	&\textit{0.000}	&\textit{0.00}	&\textit{0.000}	&	0.80 (0.47)\\	
V121	&	RRab	&	0.53518400	&	15.727	&	0.72	&	+	&	i		&	--		&--		&--		&	--		&8.65 (0.88)\\	
V123	&	RRab	&	0.54545524	&	--	&	\textit{1.10}&	--	&	$\textrm{p}_{\textrm{i}}$	&	\textit{9.55}	&\textit{0.349}	&\textit{17.51}	&\textit{0.639}	&	7.41 (0.29)\\	
V124	&	RRab	&	0.75243600	&	15.548	&	0.34	&	--	&	p		&	1.43	&0.052	&1.90	&0.069		&5.58 (0.87)\\	
V125	&	RRc		&	0.34982085	&	15.729	&	0.40	&	--	&	i		&	--		&--		&--		&	--		&1.92 (0.34)\\	
V126	&	RRc		&	0.34840456	&	15.623	&	0.40	&	--	&	s		&	0.00	&0.000	&0.00	&0.000		&1.00 (0.76)\\	
V128	&	RRc		&	0.29204160	&	15.635	&	0.53	&	--	&	s		&	0.00	&0.000	&0.00	&0.000		&1.99 (0.40)\\	
V129	&	RRc		&	0.40608800	&	15.492	&	0.40	&	--	&	n		&	-4.23	&-0.155	&-10.42	&-0.381		&6.72 (2.48)\\	
V130$^{*}$    &	RRab	&	0.56873600	&	15.690	&	0.70	&	+	&	ii		&	--		&--		&--		&--		&	--\\		
\end{tabular}
\end{minipage}
\end{table*}

\begin{table*}
\begin{minipage}{\textwidth}
\contcaption{}
\begin{tabular}{l@{\hspace{2mm}}ccccccrrrrr}
Star$^{a}$&	&	$P_a^{b,c}$		& $V_\mathrm{i}^d$		&	$A_V^{e,f}$	&Blazhko&Type$^{g}$&	\multicolumn{2}{c}{$\beta=\dot{P}$} & \multicolumn{2}{c}{$\alpha=P^{-1}\dot{P}$}  & $(P_{max}-P_{min})P_{a}^{-1}$\\
\hline
	&			&	d 			&\multicolumn{2}{c}{mag}&		&		& $10^{-10}$& $\mathrm{d Myr^{-1}}$& $\mathrm{10^{-10}d^{-1}}$& $\mathrm{Myr^{-1}}$	&$10^{-5}$\\
\hline
V131	&	RRc		&	0.29769225	&	15.684	&	0.51	&	--	&	i		&	--		&--		&--		&--		&	1.65 (0.32)\\	
V132	&	RRc		&	0.33985800	&	15.700	&	0.40	&	--	&	$\textrm{n}_{\textrm{i}}$	&	\textit{-1.78}	&\textit{-0.065}	&\textit{-5.24}	&\textit{-0.191}	&	3.53 (0.65)\\	
V133	&	RRab	&	0.55072217	&	15.686	&	0.93	&	+	&	s		&	0.00	&0.000	&0.00	&0.000		&2.27 (0.50)\\	
V135	&	RRab	&	0.56838750	&	--		&	\textit{0.75}	&	--&	s	&	0.00	&0.000	&0.00	&0.000		&4.57 (1.78)\\	
V137	&	RRab	&	0.57514720	&	15.623	&	0.90	&	--	&	p		&	0.80	&0.029	&1.39	&0.051		&2.43 (1.34)\\	
V139	&	RRab	&	0.55999450	&	15.579	&	1.21	&	--	&	n		&	-3.60	&-0.132	&-6.43	&-0.236		&2.45 (0.38)\\	
V140	&	RRc		&	0.33313425	&	15.531	&	0.44	&	+	&	i		&	--		&--		&--		&--			&3.24 (0.21)\\	
V142	&	RRab	&	0.56862700	&	15.684	&	0.95	&	--	&	p		&	0.84	&0.031	&1.48	&0.054		&4.29 (2.91)\\	
V144	&	RRab	&	0.59680200	&	15.647	&	0.62	&	+	&	i		&	--		&--		&--		&--		&	12.45 (0.54)\\	
V152	&	RRc		&	0.32612450	&	15.541	&	0.44	&	--	&	p		&	5.72	&0.209	&17.53	&0.640		&5.83 (1.66)\\	
V167	&	RRab	&	0.64397550	&	15.680	&	0.38	&	--	&	$\textrm{n}_{\textrm{i}}$	&	\textit{-10.43}	&\textit{-0.381}	&\textit{-16.20}	&\textit{-0.592}	&	7.94 (3.55)\\	
V177	&	RRc		&	0.34834442	&	15.516	&	0.54	&	--	&	s		&	0.00	&0.000	&0.00	&0.000		&3.47 (0.63)\\	
V178	&	RRc		&	0.26696282	&	15.703	&	0.36	&	--	&	s		&	0.00	&0.000	&0.00	&0.000		&3.00 (0.78)\\	
V202	&	RRab	&	0.77357100	&	15.555	&	0.15	&	--	&	s		&	0.00	&0.000	&0.00	&0.000		&3.00 (1.73)\\	
V203	&	RRc		&	0.28979525	&	15.589	&	0.15	&	--	&	s		&	0.00	&0.000	&0.00	&0.000		&0.90 (1.20)\\	
\hline
\multicolumn{12}{l}{\footnotesize{$^{a}$  No $O-C$ solutions are given for variables denoted by '*' because of their extremely complex period variations.}}\\
\multicolumn{12}{l}{\footnotesize{$^{b}$ The pulsation periods given are adopted to construct the $O-C$ plots.  }}\\
\multicolumn{12}{l}{\footnotesize{$^{c}$ The period is valid for the latest-epoch observations for stars denoted by '*'.}}\\
\multicolumn{12}{l}{\footnotesize{$^{d}$ The mean magnitude corresponds to the constant term of an appropriate order Fourier series fit of the CCD $V$ data.  }}\\
\multicolumn{12}{l}{\footnotesize{$^{e}$ Full amplitude of the mean light curves of Blazhko stars are given. }}\\
\multicolumn{12}{l}{\footnotesize{$^{f}$ No or defective CCD photometry is available for V113, V115, V123 and V135. $A_V$ is estimated from $A_{pg}$ ($A_V\approx A_B/1.3$). }}\\
\multicolumn{12}{l}{\footnotesize{$^{g}$ Period-change type as described in Sect. \ref{tablasect}; $p, s, n$ and $p_i, s_i, n_i$ denote regular and quasi-regular period-increasing, stable-period }}\\
\multicolumn{12}{l}{\footnotesize{\,\,\, and period-decreasing variables, respectively, $i$ and $ii$ are for variables showing irregular and extremely irregular period changes.}}
\end{tabular}
\end{minipage}
\end{table*}

The period-change and light-curve characteristics of the analyzed stars are summarized in Table \ref{tabla}. The first three columns list the designation, the type of the variable, and the period used for the construction of the $O-C$ diagram. 

The next two columns give photometric information on the light curves: $V_\mathrm{i}$ is the intensity-averaged mean $V$ magnitude, and $A_V$ is the full amplitude in the $V$ band. 

The sixth column in Table \ref{tabla} refers to the presence of Blazhko-type light-curve variations. Variables showing light-curve modulation indisputably in any of the available observations are denoted by `+'. Some variables with contradictory Blazhko classification and those which have not been classified as a Blazhko variable in any of the previous studies are discussed in Sect.~\ref{comments}.

The seventh column characterizes the period variation of the stars. The letters $s, p$ and $n$ indicate stable pulsation period and positive or negative period-change rates, respectively. The $O-C$ points of these stars are fitted by first- or second-order polynomials. If the coefficient of the second-order term of a parabolic $O-C$ fit is smaller than its $2\sigma$ error, the $O-C$ is fitted by a straight line and the star is labelled by $s$.

Many variables in M3 exhibit small-amplitude, irregular and/or sine-like $O-C$ variations superimposed on the steady period changes. These $O-C$s can be accurately fitted by higher-order polynomials. The slight irregularities of the period-change behaviour are indicated  by an $i$ subscript of the $s,p,n$ nomenclature for these stars. If the period change is not strictly monotonic and the period-normalised full range of the $O-C$ variation [$((O-C)_{max}-(O-C)_{min})/P$] is smaller or larger than 0.1 the classification is $s_i$ and $i$, respectively. If the $O-C$ variation indicates strictly monotonic, nonlinear period change and there is no abrupt jump in the period variations the classification is $n_i$ and $p_i$ for variables with decreasing and increasing periods, respectively. Hereafter, variables with $s_i$, $p_i$, $n_i$ period-change type are referred to as quasi-regular. The five variables (V5, V50, V79, V99, V130) showing extreme rapid and large period changes that made the construction of a real $O-C$ solution impossible are denoted by $ii$.

The the next four columns in Table \ref{tabla} give quantitative information on the observed period-change rates of the variables: 
$\beta=\dot{P}$ in $10^{-10}\textrm{dd}^{-1}$ and $\textrm{d}\textrm{Myr}^{-1}$, and $\alpha=\dot{P}P^{-1}$ in  $10^{-10} \textrm{d}^{-1}$ and $\textrm{Myr}^{-1}$ units. The period-change rates of quasi-regular stars are calculated from parabolic $O-C$ solutions; these parameters are typeset in italics. The observed, normalized full range of the period variations $(P_{max}$-$P_{min})/P$ and their errors are given in the last column.

The photometric quantities ($V_\mathrm{i}$ and $A_V$) are determined using all the available good-quality CCD $V$ data for most of the stars. Utilizing all the available observations is crucial to obtain reliable mean magnitudes, especially for variables with period close to 0.5 d in order to obtain full phase coverage, and for variables showing strong light-curve variations.
 
When zero-point offsets between the different observations are evident, the magnitudes  are shifted in order to obtain the most reliable light-curve shape and amplitude. As a rule, the CCD $V$ light curves are matched to the observations published in B06 if any exists.
The only exception is V71. The mean magnitude of the B06 $V$ light curve of V71 is 15.78 mag, which is $\sim 0.08$ mag fainter than given in any of the other observations. The B06 light curve of V71 is therefore shifted by 0.08 mag to match the other (C01, H05 and KS09) data.
CCD $V$ observations of V82, V91 and V114 have been obtained only by C01 and in the present work (for V91). These light curves are used without any magnitude shift. 

\begin{table}
\caption{Magnitude zero-point differences of the CCD $V$ light curves relative to the B06 photometry. RRab stars with stable light curves are used for the statistics.}
 \label{comp}
  \begin{tabular}{lrrrrr}
  \hline
Photometry& mean & s.d. & max & min & No.\\
\hline
C98& $ 0.006$ & $0.026$ &  $0.05$& $-0.02$& 5\\
C01& $-0.018$ & $0.023$ &  $0.04$& $-0.09$& 35\\
K98& $-0.006$ & $0.007$ &  $0.00$& $-0.02$& 14\\
H05&  $0.017$ & $0.040$ &  $0.10$& $-0.07$& 33\\
KS09&  $0.012$ & $0.025$ &  $0.09$& $-0.04$& 32\\
\hline
\end{tabular}
\end{table}
\begin{table*}
\caption{Fourier parameters of the combined CCD $V$ light curves of regular and Blazhko RRab stars. The complete table is given in the electronic version of the article as supporting information.}
\label{four}
\begin{tabular}{l@{\hspace{1mm}}c@{\hspace{2mm}}c@{\hspace{2mm}}c@{\hspace{2mm}}c@{\hspace{2mm}}c@{\hspace{2mm}}c@{\hspace{2mm}}c@{\hspace{2mm}}c@{\hspace{2mm}}c@{\hspace{2mm}}c@{\hspace{2mm}}c@{\hspace{2mm}}c@{\hspace{2mm}}c@{\hspace{2mm}}c@{\hspace{2mm}}c@{\hspace{0.5mm}}c@{\hspace{0.5mm}}c@{\hspace{0.5mm}}c@{\hspace{1mm}}l}
\multicolumn{19}{l}{RRab variables with stable light curves}\\
\hline
Star&\multicolumn{2}{c}{Type$^a$}&$P$	&$V_{m}$&$V_{i}$&$A_1$&	$A_2$&	$A_3$&	$A_4$&	$A_5$&	$\phi_{21}$&$\phi_{31}$&$\phi_{41}$&$\phi_{51}$&$A_V$&\multicolumn{3}{l}{Datasets used$^b$}\\
&$O-C$&Oo&d&\multicolumn{7}{c}{mag}&\multicolumn{4}{c}{rad}&\multicolumn{1}{c}{mag}&\\
\hline
V1&	n&	I&	0.5205962&	15.690&	15.632&	0.405&	0.175&	0.137&	0.096&	0.069&	2.341&	4.734&	1.105&	3.749&	1.17&	19, 22\\
....&....	&	....&	....&	....&	....&	....&	....&	....&	....&	....&	....&	....&	....&	....&	....&	....&\\
\\
\multicolumn{19}{l}{Blazhko RRab variables}\\
\hline
Star&\multicolumn{2}{c}{Type}&$P$	&$V_{m}$&$V_{i}$&$A_1$&	$A_2$&	$A_3$&	$A_4$&	$A_5$&	$\phi_{21}$&$\phi_{31}$&$\phi_{41}$&$\phi_{51}$&$\overline{A_V}$&$A_{min}$&$A_{max}$&$\Delta\phi$&Datasets used\\
&$O-C$&Oo&d&\multicolumn{7}{c}{mag}&\multicolumn{4}{c}{rad}&\multicolumn{3}{c}{mag}&d&\\
\hline
V3&	i&	II	&0.5581978&	15.649&	15.590&	0.408&	0.203&	0.137&	0.088&	0.062&	2.307&	4.968&	1.314&	3.993&	1.17&	1.00&	1.34&	0.008&	19, 22\\
....&....	&	....&	....&	....&	....&	....&	....&	....&	....&	....&	....&	....&	....&	....&	....&	....&	....&	....&	....\\
\hline
\multicolumn{19}{l}{\footnotesize{$^{a}$  Period-change type as introduced in Table~\ref{tabla}, and Oosterhoff classification.}}\\
\multicolumn{19}{l}{\footnotesize{$^{b}$  Numbers refer to the datasets as given in Table~\ref{ccd}.}}\\
\end{tabular}
\end{table*}

Because of the extreme changes in the amplitudes and periods of the pulsation modes and in their modulation properties, the different CCD data of double-mode/Blazhko stars V13,  V79, V99, V119 and V130 cannot be coherently combined. Therefore, only the most dense KS09 data are used to define the light-curve parameters of these stars. The differences between the mean magnitudes published in B06 and in the KS09 data are $0.01, -0.02, 0.01, 0.04$ and $-0.02$ mag for the mentioned stars, respectively.)

Zero-point differences are the most accurately determined for large-amplitude variables which do not show any light-curve modulation. Table~\ref{comp} summarizes the offsets between the B06 and the other CCD $V$ data for 51 RRab stars of stable light curve. No systematic offsets larger than 0.02 mag between the observations are evident, but the mean magnitudes of some variables may differ by $\approx0.1$ mag.

The combined, magnitude-homogenized CCD $V$ data yield very accurate light-curve parameters of the variables. The Fourier parameters of the 104 RRab variables that have good quality CCD observations are available from the electronic edition. Table~\ref{four} lists the parameters of a regular and a Blazhko RRab star as examples. The period-change and Oosterhoff types (see Sect. \ref{oosect}), the period valid for the CCD data, the magnitude- and intensity-averaged mean $V$ brightnesses, the Fourier amplitudes and the epoch-independent phase differences, the full $V$ amplitude, and the reference code of the observations combined to define the Fourier parameters (according to the numbering given in Table~\ref{ccd}) are given for each star. For Blazhko variables, the smallest and the largest observed amplitudes and the observed range of the phase oscillation of the rising branch are also given. V119 has a duplicate entry; data prior and after 2008 define its stable light-curve and Blazhko light-curve parameters, respectively.

{\begin{figure}
\includegraphics[width=8.4 cm]{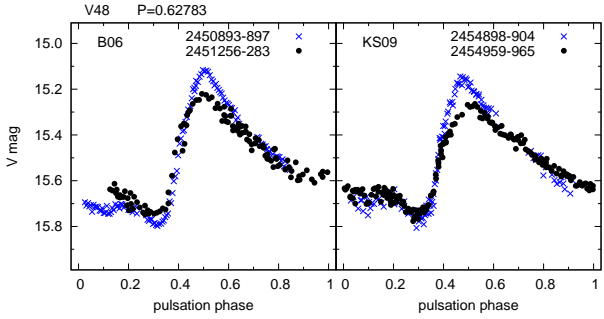}
\caption{The B06 and KS09 $V$ light curves of V48 both indicate amplitude modulation of the light curve of about $0.1$ mag.}
\label{v48}
\end{figure}
\begin{figure}
\includegraphics[width=7.7 cm]{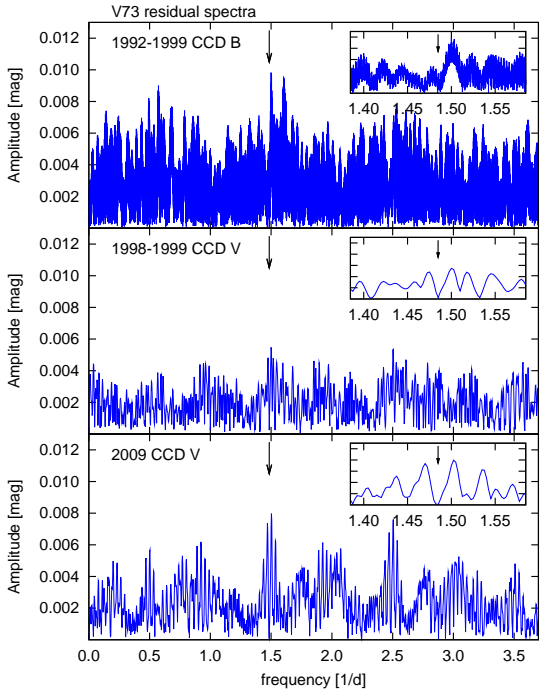}
\caption{Residual Fourier spectra of V73 for the combined C01, H05 and B06 CCD $B$ data (top panel), H05 and B06 CCD $V$ data (middle panel), and for the 2009 measurements (bottom panel). Arrows indicate the values of the pulsation frequency removed from the data. The highest residual signals appear at $\sim f_0+0.015$ in each dataset, indicating the presence of the Blazhko effect.}
\label{v73}
\end{figure}
\begin{figure}
\includegraphics[width=6.5 cm]{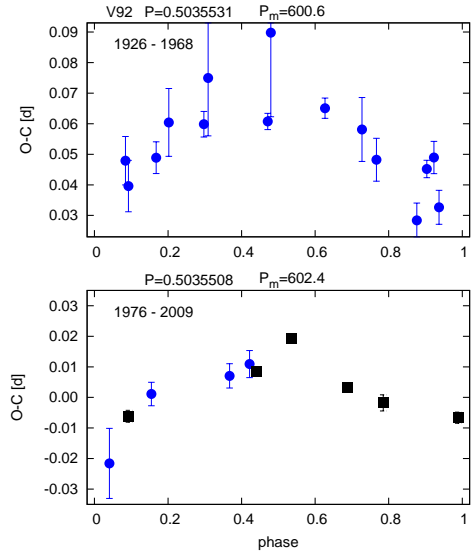}
\caption{The $O-C$ of V92 defined by yearly normal light curves of the photographic and the CCD data shows phase variation of about one hour. The $O-C$ data fit a $\sim600$ d phase oscillation  (Blazhko modulation) for the 1926--1968 and the 1976--2009 intervals as shown in the figure. Filled circles and squares are for photographic and CCD observations, respectively. }
\label{v92}
\end{figure}
\begin{figure}
\includegraphics[width=8.4 cm]{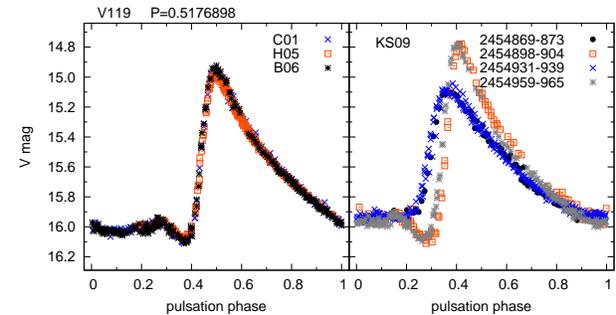}
\caption{CCD $V$ light curves of V119.
No light-curve modulation is evident in the combined CCD observations obtained between 1992 and 1999 as shown in the left-side panel. In the right-hand panel, the 2009 observations show very strong amplitude and phase modulations with a Blazhko period of 50--60 days. Most of the photographic data showed stable light curves, however, \citet{la22} noted anomalous shapes of maxima, indicating that the light curve of V119 might have also been modulated at that time.}
\label{v119}
\end{figure}
\begin{figure}
\includegraphics[width=8.4 cm]{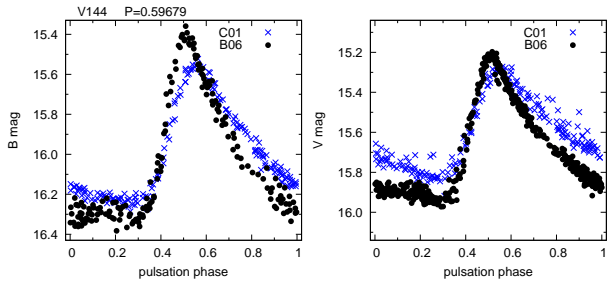}
\caption{Comparison of the C01 and B06 CCD $B$ and $V$ light curves of V144. The 0.15--0.20 mag amplitude difference between the two datasets indicates light-curve  modulation with a long period. Although the mean magnitudes of the C01 $B$ and $V$ light curves are $\sim0.1$ mag brighter than in the B06 data, the bright maxima of the B06 light curves contradicts that the difference between the light curves originates from crowding.}
\label{v144}
\end{figure}
\begin{figure}
\includegraphics[width=6. cm]{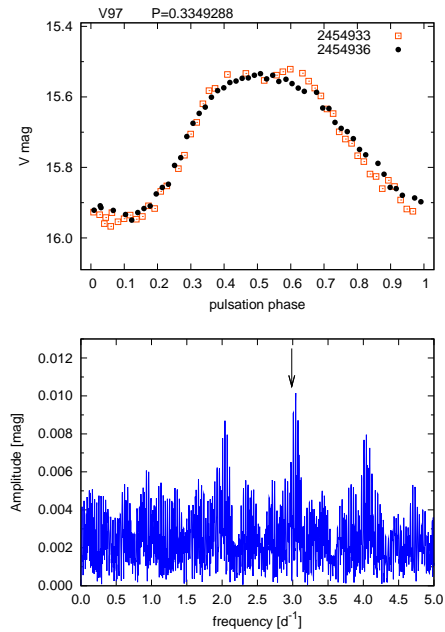}
\caption{Light-curve variations in the height and shape of the maxima of V97 are detected as shown in the top panel. The Fourier spectrum of the residual KS09 data (bottom panel) shows a frequency component at $f_0+f_m$ ($f_m=0.056 \textrm{ cd}^{-1}$; $P_m=17.7\textrm{ d}$). The arrow points to the position the pulsation frequency removed.}
\label{v97}
\end{figure}
\begin{figure}
\includegraphics[width=6.2 cm]{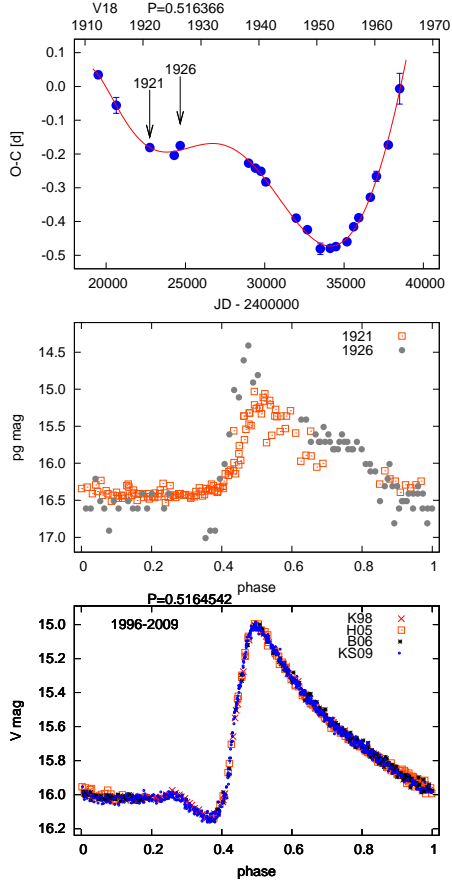}
\caption{Seasonal $O-C$ variation during the first half of the last century (top panel) and the folded light curves of V18 in two different years (middle panel). The arrows on the top panel mark the yearly datasets plotted. The large variation of the pulsation period makes the analysis of the Blazhko effect impossible, but the changes of the pulsation amplitude indicate light-curve modulation undoubtedly. The CCD data obtained between 1996 and 2009 (bottom panel) show stable pulsation light curve.}
\label{v18}
\end{figure}
\begin{figure}
\includegraphics[width=6.2 cm]{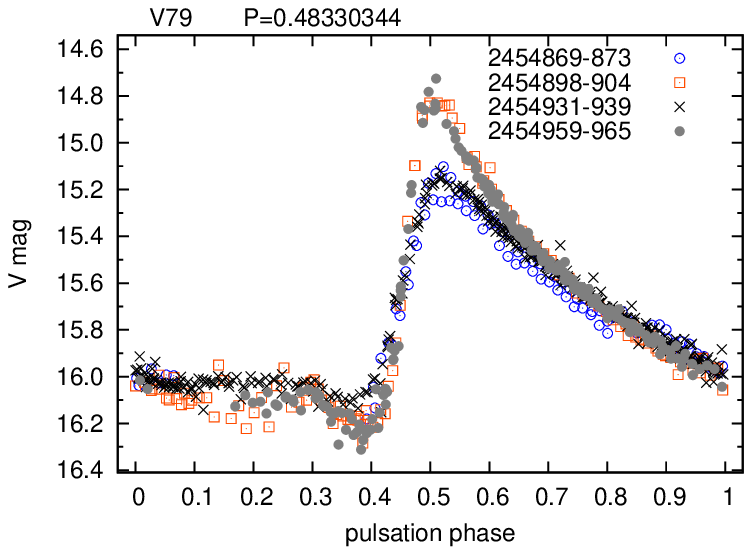}
\caption{The KS09 $V$ light curve of V79 from 2009. It is a typical RRab star now, showing strong amplitude modulation. No sign of the first overtone mode is detectable.}
\label{v79}
\end{figure}}

\subsubsection{New and temporal Blazhko variables}
\label{comments}
Blazhko-type light-curve modulation of five RRab stars (V48, V73, V92, V119 and V144) and one  RRc variable (V97) has been revealed for the first time from the 2009 observations or has become evident with the combination of the different CCD data, as documented in Figs.~\ref{v48},~\ref{v73},~\ref{v92},~\ref{v119},~\ref{v144}, and \ref{v97}. No  light-curve variability of these stars has been reported previously, either because it has escaped detection or because of the recent onset of the modulation.

The photographic light curve of V18 showed large amplitude modulation and irregular period changes in the first half of the last century (see Fig. ~\ref{v18}).  The Blazhko modulation of the star ceased after a significant pulsation-period increase in 1951. None of the CCD observations shows evidence of light-curve variability of V18 in the last decades either. A similar behaviour, -- disappearence of the modulation after a strong period increase --, was detected in a field RR Lyrae star, RR Gem \citep{rrg}, too.

The unique behaviour of V79 was discussed by \citet{cl99} and \citet{go10} in detail. The photographic observations showed large irregular period variations and amplitude changes of the light curve. The CCD observations revealed double-mode nature of V79 dominated by the first-overtone and the fundamental-mode pulsations between 1992--2006 and in 2007, respectively. According to the latest observations (after 2008), the star is a fundamental mode variable again showing large-amplitude Blazhko modulation as documented in Fig.~\ref{v79}. 

\section{Does stellar evolution explain the observed period changes?}

Theory predicts that the evolutionary tracks of RR Lyrae stars, starting from the Zero Age Horizontal Branch (ZAHB) run first blueward, then redward at slightly higher luminosity  \citep{sr79,ld,dor,pi04}. According to canonical models, the blueward evolution is relatively slow, resulting in a small negative rate of period change ($\alpha  \gtrsim -2\times 10^{-10}\textrm{d}^{-1}$). As the star evolves from blue to red, toward the end of core helium burning, the period increases at a higher rate ($\alpha  \approx$ 1--10 $ \times 10^{-10}\textrm{d}^{-1}$). The stars are at  $\approx$ 0.05--0.10 mag brighter mean $V$ magnitudes in this evolutionary stage than on the ZAHB. The evolutionary tracks of variables, with ZAHB position bluer than the instability strip, intersect the instability strip at  0.10--0.20 mag above the ZAHB. The period-increase rate of these very few variables can be  as high as $\alpha \approx$ 10--30 $\times 10^{-10}\textrm{d}^{-1}$.

The major goal of this study is to investigate if the observed period changes of M3 RR Lyrae stars are consistent, at least in a statistical sense, with theoretical predictions. 
\begin{figure}
\includegraphics[width=8.2 cm]{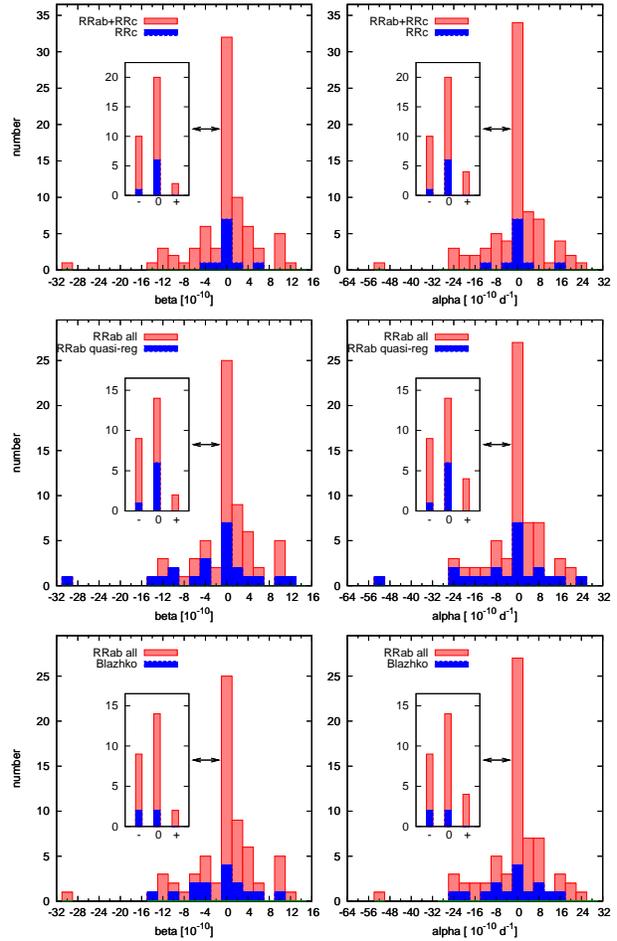}
\caption{Histograms of the period-change rates of RR Lyrae stars in M3 with regular and quasi-regular period-change properties. The inserts show the fine structure of the period-change distributions of the most populated bin around zero. The left- and right-hand panels show the distributions of $\beta$ $[10^{-10}]$ and $\alpha$ $[10^{-10}\textrm{d}^{-1}]$, respectively. On the top panels, the distributions of the period-change rates for all the studied variables are documented. RRc/RRd stars are shown by the dark areas. The histograms in the middle and bottom panels display the results only for RRab stars. Dark areas correspond to quasi-regular period changes and Blazhko variables in the middle and bottom panels, respectively. The histograms indicate very slight, if any, dominance of positive period-change rates. The surplus of variables with positive period-change rates is partly balanced by the large number of stars with small period-decrease rates in the central bins, as shown in the inserts.}
\label{histogram}
\end{figure}
\begin{figure}
\includegraphics[width=8.0 cm]{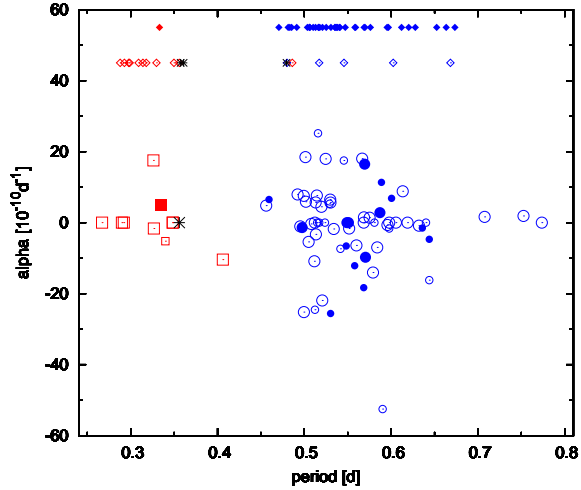}
\caption{Normalized period-change rate ($\alpha$) versus period plot of M3 RR Lyrae stars. Circles and squares shown by large and small symbols denote RRab and RRc stars with regular period variations  ($s, p, n$) and with small irregularities superimposed  ($s_i, p_i, n_i$), respectively. Blazhko stars are shown by filled symbols. To indicate the period distribution of the whole sample,  stars with irregular period changes ($i, ii$) are set at the arbitrary values of 45 (regular light curves) and 55 (modulated light curves). Double-mode variables are denoted by asterisks. Positive and negative period-change values are equally frequent in the sample, indicating that the mean period-change rate of the variables is very close to zero. Large period changes (both negative and positive values) tend to occur in shorter-period RRab stars, while the period-change rates of longer-period variables are modest. }
\label{p-alfa}
\end{figure}
\begin{figure}
\includegraphics[width=8.0 cm]{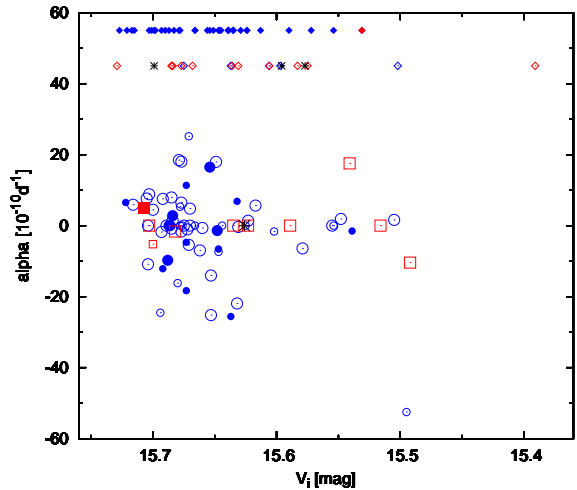}
\caption{Normalized period-change rate ($\alpha$) versus mean $V_\mathrm{i}$ magnitude. The symbols are the same as in Fig. \ref{p-alfa}.}
\label{v-a-alfa}
\end{figure}

\subsection{On the distribution of the period-change rates}

Fig.~\ref{histogram} shows the histograms of the period-change rates for the studied RR Lyrae sample and its different subsamples. The inserts resolve the most populated bin around zero, and show that variables of small negative period-change rates are more numerous than variables of similarly small period-increase rates in each sample. The period-change rates of these stars fit fairly well the predictions of the models for the blueward phase of the HB evolution, i.e., evolutionary period changes are probably directly observed in these stars.  The distribution of the mean-brightness values of these variables confirms this possibility. Seven of the nine RRab variables that show small period-decrease rates have mean-brightness value in the 15.63--15.69 mag range. On this account, these stars are only slightly, if at all above the  ZAHB position of M3. The two stars at brighter luminosities are V14 and V26. V14 is a Blazhko variable, therefore its period variation does not necessarily reflect evolutionary changes. The bright luminosity of V26 indicates that it is in a more advanced evolutionary state. 

The normalized period-change rate ($\alpha$) versus period and mean-magnitude plots are shown in Figs. \ref{p-alfa} and \ref{v-a-alfa}. Positive and negative period-change rates with similar size  are equally frequent at any period and brightness value. In the sample of variables with regular and quasi-regular period changes, there are 8 stars with large period-increase rates ($\alpha > 10 \times 10^{-10}\textrm{d}^{-1}$). Such rapid evolution is allowed  by theory only for a very small sample of overluminous variables.
However, the mean $V$ magnitudes place these stars close to the ZAHB (see Fig~\ref{v-a-alfa}) that contradicts their evolved state.
There is only one bright RRc star (V152) among these variables with  $V=15.55$ mag mean brightness, 0.15 mag above the ZAHB.
It is also worth noting that none of the four brightest RR Lyr has an increasing period:\\
V70: $V=15.391$, irreg. period change, \\
V129: $V=15.492, \alpha =-10.4 \times 10^{-10}\textrm{d}^{-1}$,\\
V42: $V=15.495, \alpha =-52.5 \times 10^{-10}\textrm{d}^{-1}$, and \\
V65: $V=15.495$, irreg. period change.

As to the stars with decreasing periods, not only the extreme period-change rate of V42 ($\alpha = -52 \times 10^{-10}\textrm{d}^{-1}$), but that of the vast majority of these variables cannot be interpreted by standard evolutionary models.  

Recently, \citet{si10} modelled the period-change behaviour of pre-ZAHB stars in M3 and concluded that about two pre-ZAHB variables with large period-decrease rates should be present in the total sample. This is in high contrast with the 20 observed period decreasing RR Lyrae stars with steady or quasi-regular period change rates of $\alpha < -2 \times10^{-10}\textrm{d}^{-1}$.

\begin{table}
\caption{Statistics of the period-change rates.}
\label{stat}
\begin{tabular}{llllr}
\hline
	&		&$\overline{\beta}$ ($\textrm{d}$ $\textrm{Myr}^{-1}$)	&$\overline{\alpha}$ ($\textrm{Myr}^{-1}$)&	No.\\
\hline
\multicolumn{5}{l}{Total sample ($s, n, p, s_i, n_i, p_i$)}\\
RRc	&mean	&$+0.003$ $\pm$ $0.087$	&$+0.018$ $\pm$ $0.250$	&11\\
	&m.p.v.$^a$	&$+0.000$	&$-0.012$	&\\
RRab&mean	&$-0.022$ $\pm$ $0.243$	&$-0.036$ $\pm$ $0.439$	&66\\
	&m.p.v.	&$+0.000$	&$+0.003$	&\\
\hline
all	&mean	&$-0.018$ $\pm$ $0.227$	&$-0.028$ $\pm$ $0.417$	&77\\
	&m.p.v.	&$+0.002$	&$+0.004$ &\\
\hline
\hline
\multicolumn{5}{l}{Variables of strictly linear period changes ($s, n, p$)}\\
RRc	&mean	&$+0.010$ $\pm$ $0.089$	&$+0.039$ $\pm$ $0.254$	&10\\
	&m.p.v.	&$+0.000$	&$+0.000$	&\\
RRab&mean	&$+0.015$ $\pm$ $0.172$	&$+0.026$ $\pm$ $0.323$	&44\\
	&m.p.v.	&$+0.019$	&$+0.036$	&\\
\hline
all	&mean	&$+0.014$ $\pm$ $0.159$	&$+0.029$ $\pm$ $0.309$	&54\\
	&m.p.v.	&$+0.016$	&$+0.032$	&\\
\hline
\hline
\multicolumn{5}{l}{Without Blazhko stars and the one outlier (V42) }\\
RRc &mean	&$-0.003$ $\pm$ $0.090$	&$+0.001$ $\pm$ $0.257$	&10\\
	&m.p.v.	&$+0.000$	&$+0.000$	&\\
RRab&mean	&$+0.008$ $\pm$ $0.194$	&$+0.017$ $\pm$ $0.365$	&50\\
	&m.p.v.	&$+0.015$	&$+0.029$	&\\
\hline
all	&mean	&$+0.006$ $\pm$ $0.180$	&$+0.014$ $\pm$ $0.347$	&60\\
	&m.p.v.	&$+0.011$	&$+0.018$	&\\
\hline
\hline
\multicolumn{5}{l}{\footnotesize{$^{a}$ most probable value; mode}}\\
\end{tabular}
\end{table}

Table~\ref{stat} summarizes the statistics of period changes of RR Lyrae stars with regular and quasi-regular period changes in M3. The median value of the period-change rates equals zero for the total sample and for any of the subsamples. Small, either positive or negative rates are derived for the total sample depending on whether the mean or the most-probable values are calculated, and $\alpha$ or $\beta$ period-change rates are used for the statistics. If variables with the most regular behaviour are only considered (stars with strictly linear period changes or variables not showing the Blazhko effect, and omitting the most extreme period decrease value of V42), the results converge to a small positive, $\beta \approx\, 0.01\, \textrm{d} \textrm{Myr}^{-1}$  value. 

\begin{figure*}
\includegraphics[width=15.6cm]{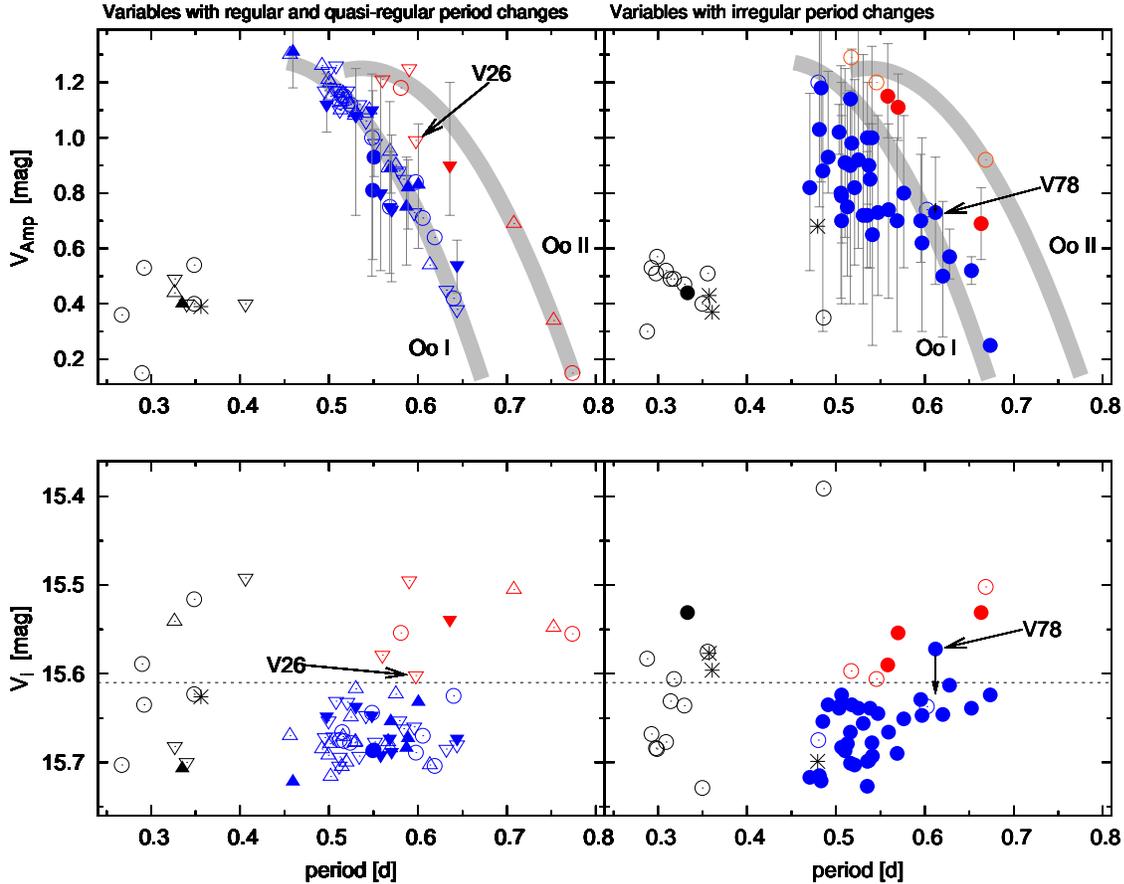}
\caption{Period--amplitude ($P-A$) and period--intensity-averaged $V_\mathrm{i}$ magnitude ($P-V_\mathrm{i}$) diagrams of M3 variables.  Variables with regular and quasi-regular period changes are shown in the left-hand panels. Circles denote stars with stable period, upward and downward triangles indicate increasing and decreasing periods, respectively. No distinction is made between regular and quasi-regular period changes. Variables showing irregular period variations are plotted in the right-hand panels. Blazhko stars are denoted by filled symbols in every panel. The full range of the  observed amplitude variations of Blazhko stars are indicated by vertical lines in the $P-A$ planes.  Double-mode variables are denoted by asterisks. The $P-A$ relations of OoI and OoII variables defined by 42 and 9 stable light-curve RRab stars (open symbols), are drawn by grey strips in the top panels. The OoI and OoII samples are separated at $\sim15.61$ mag brightness as shown in the  $P-V_\mathrm{i}$ planes. Two stars are labelled in the figures. The brightness and the amplitude of V26 both indicate an evolved status of the star (see in Sect 5.1).  The bright mean magnitude of V78 is the consequence of the contamination of a close  companion, the real magnitude of the star is shown by the vertical arrow (see in Sect. 5.2). For a colour version of the figure see the online edition of the paper.}
\label{p-ampv}
\end{figure*}

\subsection{Period changes on the Oosterhoff sequences}
\label{oosect}
The period-change studies of RR Lyrae stars have revealed that the period-change behaviour of stars of a GC is related to its Oosterhoff (Oo) type \citep{rs97}. The tendency is that both the mean and median values of $\alpha$ (or $\beta$) are smaller for OoI  than for OoII clusters in keeping with evolutionary models; most of the RR Lyrae stars in OoI clusters are located near their ZAHB position while most of those in OoII clusters have already left their initial position on the HB and are moving from blue to red through the instability strip toward the asymptotic red giant branch \citep{ldz}, producing a surplus of increasing periods. Consequently, one would expect that the RR Lyrae stars on the densely populated OoI sequence of M3 differ remarkably from the few OoII-sequence stars as to their period changes. To check the validity of this scenario, we have investigated the period changes on the Oo sequences separately, too.

The period--amplitude ($P-A$) and period--intensity-averaged mean $V_\mathrm{i}$ magnitude ($P-V_\mathrm{i}$) diagrams of M3 variables are shown in Fig. \ref{p-ampv}. The left-hand and right-hand panels show these plots for variables with regular and quasi-regular period changes and irregular period-change properties, respectively. The Oo sequences defined by non-Blazhko variables are also drawn in the $P-A$ planes. (Blazhko stars are not used to define the Oo sequences because of their uncertain amplitudes.) 

Among the 42 and 9 RRab stars tracing the OoI and OoII sequences, 9, 16, 15, 2 and 2, 2, 2, 3  stars have stable, increasing, decreasing and irregular period-change properties, respectively. Each period-change property seems to be equally probable on both Oo sequences. (The low number of irregular period-change type on the OoI sequence is the consequence of the fact that most of the variables showing irregular period changes exhibit the Blazhko effect and Blazhko stars were not used to define the Oo sequences). A similar result has recently been obtained by \cite{ic4499}, studying the period-change properties of RR Lyrae stars in IC4499. 

The selection of stars defining the OoI and OoII sequences is based on their positions on the $P-A$ plane, regardless of their mean magnitudes. The observed brightness of the variables, however, fully supports our selection as shown in the bottom panels of Fig~\ref{p-ampv}. Among the variables with regular and quasi-regular period-change properties shown in the left-hand panels, each RRab star brighter than 15.61 mag (drawn by dashed line in the bottom panels) is above the Oo I sequence. With one exception, this is also true for variables with irregular period changes shown in the right-hand panels. The exception is V78, a Blazhko variable with a pulsation amplitude corresponding to the OoI sequence but with relatively bright, $V_\mathrm{i}=15.572$ mag mean brightness. The inspection of SDSS images \citep{sdss} has revealed that V78 has a companion at 0.09" separation with $r=18.495$ mag brightness \citep{sdss2}. Most probably, this companion was not resolved in any of the used photometries.
If the magnitude of V78 is corrected for the contamination of the companion, the mean magnitude is about 0.06 mag fainter, in agreement with the OoI classification of the stars based on its amplitude (see the right panels of Fig. \ref{p-ampv}). The $\sim 0.04$ mag decrease of the amplitude caused by the companion does not influence the Oo classification.

The period-change rates of 40 non-Blazhko variables with regular and quasi-regular period variations on the OoI sequence are in the  $-25 <\alpha < 25\times10^{-10}\textrm{d}^{-1}$ range. The mean brightness of these stars shows that they are close to or only slightly above the ZAHB. The period-increase rates of 5, and the period-decrease rates of 10 variables are larger than expected in any steady phase of the HB evolution in this brightness range. The large number of these stars contradicts the assumption that they all are in a special, extremely rapid evolutionary phase.

Although the sample is small, the period changes of variables on the OoII sequence are not in accordance with evolutionary model predictions, either. The brightness of variables above the OoI sequence supports that these stars are indeed in the late, redward phase of the HB evolution. There are 7 non-Blazhko RRab stars with regular period-change property near or on the OoII sequence. Only two of these stars (V60 and V124) show period increase, however, with quite modest, $1.6$ and  $1.9 \times 10^{-10}\textrm{d}^{-1}$  $\alpha$ values, respectively. The modest period-increase rates of these stars, -- if they have evolutionary origin --, warn that even in the late stages of the HB evolution the period increase does not necessarily have an extremely large value. The period-change behaviour of the 3 non-modulated RRab stars above the OoI sequence with decreasing periods are in conflict with standard evolutionary explanation.

\subsection{Irregular period changes and the Blazhko effect}

The period changes of 42 of the studied 108 RRab stars are irregular. Most of these stars are Blazhko variables. 

Table~\ref{statirreg} gives the Blazhko star statistics of the studied sample of M3 RRab stars and its subsamples selected according to the period-change properties. About 50\% of the total sample of RRab stars show the Blazhko effect. The occurrence rate of the modulation is as high as 88\% among variables with irregular period variations, while it is only 16\% for variables showing linear period changes. The incidence rate of the modulation is  significantly larger among variables with quasi-regular period changes (36\%), as well. It seems that even slight irregularities are in connection with the  occurrence of the Blazhko modulation. The Blazhko statistics of RRab stars of constant and increasing periods are also compared with that of variables of decreasing periods. Again, the difference is significant. While  18\% of the variables with stable or increasing periods show the Blazhko effect, the incidence rate is 30\% among  variables with decresing periods. Moreover, the period-decrease rates of six from the eight Blazhko variables with decreasing periods are too large to be explained by  standard HB evolution.

Based on these tendencies, one may speculate whether strong and slight irregularities of the period changes, and large period decrease rates are connected to the Blazhko phenomenon. The unique, temporal appearence of the light-curve modulation of V18 and V119 warns that quite a number of other Blazhko variables may exist in M3, which just do not show light-curve variations at the snapshots of the CCD observations, and their photographic data are not accurate enough to detect the modulation. We do not know either,  how long observations are needed to reveal the true period-change behaviour of the variables. For about 90 years, V24 had not shown any notable period change, while V106 had exhibited quasi-monotonic period decrease. Then the period of both Blazhko stars changed abruptly. Based on the 110 years long dataset, the period-change type of these  variables is irregular.

All these properties warn that one has to be cautious when interpreting the period changes of RR Lyrae stars on evolutionary ground. 
One may hope to detect evolutionary changes only in  variables with strictly regular behaviour on a very long time base. A rigorous selection of Blazhko stars and variables with irregular period changes is very important to get a reliable result.

\begin{table}
\caption{Occurrence rates of the Blazhko modulation in different samples}
\label{statirreg}
\begin{tabular}{llll}
\hline
Period-change&\multicolumn{2}{c}{Number of RRab stars }&Percentage  \\
type                  &all& Blazhko&of Blazhko stars\\ 
\hline                    
all                             &108           &52     &48\\
\hline
$s,p,n$				&44		&7	&16\\
$s_i,p_i,n_i$		        &22		&8	&36\\
$i,ii$				&42		&37	&88\\
\hline
$s,s_i,p,p_i$		        &39		&7	&18\\
$n,n_i$				&27		&8	&30\\
\hline
\hline
\end{tabular}
\end{table}

\section{Does period change influence the light-curve shapes of RRab stars? }

\begin{figure}
\includegraphics[width=7.6 cm]{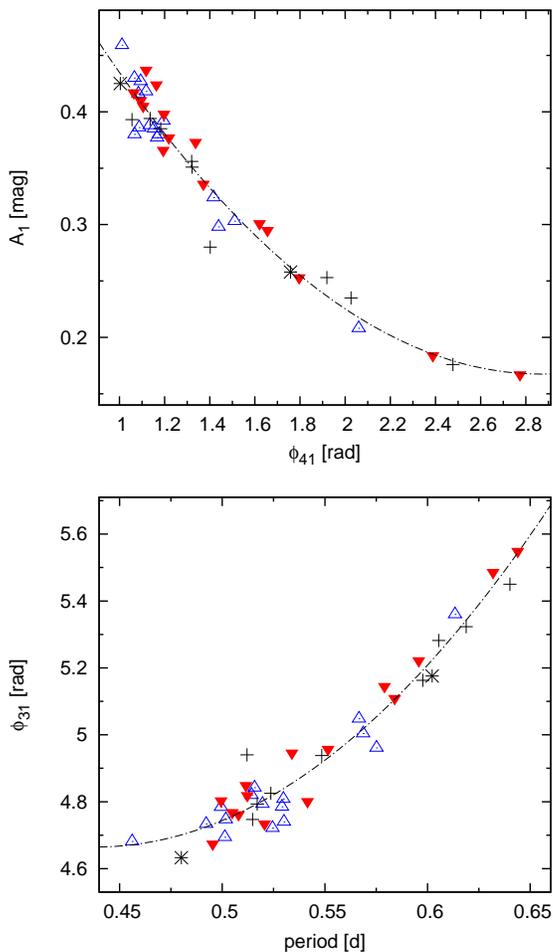}
\caption{Fourier amplitude ($\textrm{A}_1$) versus Fourier phase difference ($\phi_{41}$) and  $\phi_{31}$ versus period plots of 42, non-Blazhko, OoI type RRab stars. Stars with increasing and decreasing periods are marked by upward and downward triangles, respectively. The plus marks and asterisks denote variables with stable periods and irregular period changes, respectively. The second-order polynomial fits of the data are also shown in the figures. There is a slight tendency that the distributions of stars with increasing and decreasing periods are systematically different. In all similar planes, period increasing and decreasing variables tend to cluster below and above the fitted curves, respectively. Monte Carlo simulations confirm that this effect is statistically significant. For a colour version of the figure see the online edition of the paper.}
\label{pl}
\end{figure}

The OoI sequence of M3 is defined in our sample by 42 variables, which all have good-quality CCD light curves with accurate Fourier parameters. There are 16 and 15  stars with increasing and decreasing pulsation periods in this sample, respectively. The distributions of these stars do not show any systematic differences either on the $P-A$ plane or on any of the $P-$Fourier amplitude ($P-A_i$) plots. Consequently, they seem to belong to a very homogeneous population. These data enable us to examine whether the period-change behaviour has any effect on the light-curve shapes of the variables.

A visual inspection of the locations of variables with increasing and decreasing periods on the plots of the possible combinations of the Fourier parameters suggests that there may be some slight systematic difference between the $A_i-\varphi_{k1}$ and $P-\varphi_{k1}$ relations of variables of increasing and decreasing periods as illustrated in Fig.~\ref{pl}. 

A Monte Carlo (MC) simulation confirms the reality of the differences undoubtedly. The mean distance of the 16 period-increasing and 15 period-decreasing variables from a second-order fit of the total sample of 42 OoI stars are calculated in the different $A_i - \varphi_{k1}$, and $P-\varphi_{k1}$ planes. In all the $A_i - \varphi_{k1}$ planes, period increasing variables tend to have 0.002--0.014 mag lower amplitudes on the averages at a given phase-difference value than period decreasing variables have, while their phase differences tend to be smaller by 0.03--0.07 rad at a given period than the phase differences of variables with decreasing periods. Although the differences between the mean distances of the two groups from the common fits are smaller than $\sim 2 \sigma$ in each case, MC simulations show that it has only 0.1\% probability that in each $A_i - \varphi_{k1}$, and $P-\varphi_{k1}$ plane a mean offset between the two groups similar to or larger than what is observed occurs by chance.

The mean $V_\mathrm{i}$ brightness of the period-decreasing sample is 0.01 mag brighter that the mean brightness of the period-increasing one. Therefore, the question arises whether the observed differences between the light-curve shapes are the consequence of a slight difference in the luminosities of the two samples. In order to test this possibility, a MC analysis has also been carried out excluding the brightest (V11) and the faintest stars (V76) from the period decreasing and increasing samples, respectively. The difference between the mean $V_\mathrm{i}$ brightnesses of the two samples has been reduced to 0.004 mag in this case. It has, however, marginal effect on the results. The observed differences between the Fourier parameters of the remaining 15 and 14 stars are found to occur by chance in 0.2\% of the MC simulations. The offsets between the Fourier parameters of the period increasing and decreasing samples have not changed significantly by the omission of these two stars. The differences between the mean distances of the two samples from the common fits in the different Fourier amplitude--phase difference combinations decreased on the average by $\sim 10$ per cent, but they increased by 7 per cent for the period--phase difference combinations.

A similar test was performed by excluding the shortest- and the longest-period stars with increasing and decreasing periods, respectively. No significant differences of the results has been found in this case either.

\section{Summary and discussion}

The period-change behaviour of 134 RR Lyrae stars in M3 is investigated using observations covering a time base longer than a century. The statistics of the period-change rates yield small positive mean and most-probable period-change rates for regular subsamples:
variables showing strictly linear period changes (54 stars); variables with regular and quasi-regular period-change properties, omitting Blazhko stars and the most extreme period-decrease rate of V42 (60 stars). The $\beta \approx 0.01 \mbox{ d} \textrm{Myr}^{-1}$
mean period-increase rates derived for these subsamples agree well with theoretical predictions \citep{le91}.

The model simulations of the period-change distributions of RR Lyrae stars in GCs rely on the assumption that the observed period change of each variable reflects the actual evolutionary state of the star. In reality, however, this is not always true. 

In M3, variables showing small period-decrease rates at mean-brightness values close to or slightly above the ZAHB magnitude, and the overluminous, long-period variables with period-increase rates of $\alpha \approx 1-2 \times 10^{-10}\textrm{d}^{-1}$ on the OoII sequence reflect evolutionary period changes very likely. Nevertheless, when deriving any statistical value from the observed period-change rates, a number of variables are included, which do not fit in any phase of the HB evolution according to their period-change rate and brightness. As these variables typically have large period-change rates, they dominate the results of the statistics. Therefore, a direct comparison of the statistical results and model predictions is somewhat dubious. 

The mean $V_\mathrm{i}$ magnitudes of M3 variables belonging to the OoII sequence in the $P-A$ plane are 0.10--0.15 mag brighter than the mean magnitudes of the OoI variables. The brighter luminosity of variables near or on the OoII sequence supports the more evolved state of these stars. The period changes of many variables on the OoI and OoII sequences are, however, in conflict with theoretical expectations; either the observed period changes do not reflect the evolutionary state of the star, or theory does not interpret the HB evolution accurately enough. 

The large number of variables displaying irregular period change and/or the Blazhko effect seems to favour the former possibility.
It is worth mentioning that these stars are even more numerous in M3 than in M5 \citep{m5,m5bl}. In M3, 42 stars have irregular period changes and 52 stars show the Blazhko effect among the 108 RRab variables studied. The occurrence of the Blazhko effect and irregular period changes are strongly correlated both in M3 and M5; most of the Blazhko stars have irregular period changes and variables with unstable pulsation period often display light-curve variability. Although there is no satisfactory explanation either for the irregular period changes or the Blazhko effect, the evolutionary period changes of variables showing these irregularities are presumably masked by other effects.

We have also shown that, in some cases, the light-curve modulation appears only temporarily, and the irregularity of the period variations are revealed only after a very long time interval (more than 90 years) of regular or quasi-regular period-change behaviour. 
Therefore, it can be supposed that there is still an unidentified sample of Blazhko stars and/or variables of irregular period changes among variables showing regular behaviour based on the available observations. Some of the observed regular or quasi-regular  period-change rates that do not fit the supposed evolutionary status can probably be explained by the temporal regularity of the light curve and/or the period change behaviour of these stars.

One of the most important results of the present study is the detection of small, systematic differences between the light-curve parameters of period-increasing and period-decreasing variables. The OoI sequence of M3 is well populated and it supplies a homogeneous sample of variables with accurate light-curve parameters to perform a thorough investigation. At a given period and amplitude, the epoch-independent phase differences of the light curves of period-increasing stars tend to be slightly smaller than the $\varphi_{k1}$ values of period-decreasing stars are. Very probably, the period changes of many of the variables included in this comparison reflect other than evolutionary effects. Consequently, the slight differences between their light curves are not caused by the different evolutionary stages in the two samples. We have already shown for some field Blazhko stars that rapid period changes are accompanied by changes of the modulation properties and the shape (amplitude) of the mean pulsation light curve \citep{rrg, kbs1, czl}. Further detailed investigations of the connections between period changes and variations/differences in the light curves of RRab stars may help to answer the question, what kind of structural changes/differences  are responsible for the period-change behaviour of RRab stars.
\section*{Acknowledgments}
We thank the referee, Christine Clement, for the thorough reading of the manuscript, and for her comments.
We also thank C. Pilachowski for kindly providing us the information on the Kitt Peak observations.
The financial support of OTKA grants K-68626 and T-81421 is acknowledged. 
 
\end{document}